\pdfoutput=1

\documentclass[
aps,
prb,
amsmath,amssymb,
10pt,
twocolumn,
superscriptaddress,
]{revtex4-2}

\usepackage{graphicx}%
\usepackage{dcolumn}%
\usepackage{bm}%

\usepackage[normalem]{ulem}
\usepackage[svgnames]{xcolor} %
\usepackage[colorlinks,citecolor=DarkGreen,linkcolor=FireBrick,linktocpage,unicode]{hyperref} %
\usepackage{textcomp}  
\usepackage{rotating} 
\usepackage{comment}
\usepackage{amssymb}
\usepackage[flushleft]{threeparttable}

\begin{document}

\preprint{APS/123-QED}

\title{Global structure searches under varying temperatures and pressures using polynomial machine learning potentials: A case study on silicon}

\author{Hayato \surname{Wakai}}
\email{wakai@cms.mtl.kyoto-u.ac.jp}
\affiliation{Department of Materials Science and Engineering, Kyoto University, Kyoto 606-8501, Japan}
\author{Atsuto \surname{Seko}}
\email{seko@cms.mtl.kyoto-u.ac.jp}
\affiliation{Department of Materials Science and Engineering, Kyoto University, Kyoto 606-8501, Japan}
\author{Isao \surname{Tanaka}}
\affiliation{Department of Materials Science and Engineering, Kyoto University, Kyoto 606-8501, Japan}
\date{\today}

\begin{abstract}
Polynomial machine learning potentials (MLPs) based on polynomial rotational invariants have been systematically developed for various systems and applied to efficiently predict crystal structures.
In this study, we propose a robust methodology founded on polynomial MLPs to comprehensively enumerate crystal structures under high-pressure conditions and to evaluate their phase stability at finite temperatures.
The proposed approach involves constructing polynomial MLPs with high predictive accuracy across a broad range of pressures, conducting reliable global structure searches, and performing exhaustive self-consistent phonon calculations. 
We demonstrate the effectiveness of this approach by examining elemental silicon at pressures up to 100 GPa and temperatures up to 1000 K, revealing stable phases across these conditions.
The framework established in this study offers a powerful strategy for predicting crystal structures and phase stability under high-pressure and finite-temperature conditions.

\end{abstract}

\maketitle

\section{Introduction}

Many elemental systems and compounds are known to exhibit polymorphism, wherein they can crystallize into multiple distinct crystal structures. 
Notably, numerous polymorphs have been observed, especially under high-pressure conditions. 
To predict such crystal structures under varying pressure conditions, ranging from ambient to high pressures, global crystal structure searches using density functional theory (DFT) calculations have proven to be a powerful and effective tool \cite{Oganov2009, Pickard_2011, PhysRevLett.106.015503, doi:10.1126/science.1244989}. 
Recent studies have also shown that machine learning potentials (MLPs)
\cite{
Lorenz2004210,
behler2007generalized,
behler2011atom,
PhysRevB.92.045131,
ARTRITH2016135,
PhysRevB.96.014112,
PhysRevB.97.094106,
han2017deep,
LI2020100181,
bartok2010gaussian,
PhysRevB.90.104108,
PhysRevLett.114.096405,
PhysRevB.95.214302,
PhysRevX.8.041048,
PhysRevB.90.024101,
PhysRevB.92.054113,
Thompson2015316,
doi-10.1137-15M1054183,
PhysRevMaterials.1.043603,
PhysRevMaterials.1.063801,
wood2018extending,
https://doi.org/10.1002/qua.24836,
KHORSHIDI2016310,
10.1063/5.0158710,
doi:10.1063/1.5126336,
Freitas2022}
significantly accelerate global structure searches by facilitating efficient energy and force calculations during local geometry optimizations \cite{PhysRevLett.120.156001,PhysRevB.99.064114,GUBAEV2019148,Kharabadze2022,PhysRevB.106.014102,HayatoWakai202323053,PhysRevB.110.224102,seko2024polynomialmachinelearningpotential}. 
MLPs and their frameworks have been developed to perform large-scale simulations and a vast number of systematic calculations accurately and efficiently, which would be computationally expensive using DFT calculations.
These MLPs represent interatomic interactions by utilizing a range of structural features that describe the neighboring atomic distribution, in conjunction with machine learning models such as artificial neural networks \cite{Lorenz2004210, behler2007generalized, behler2011atom, PhysRevB.92.045131, ARTRITH2016135, PhysRevB.96.014112, PhysRevB.97.094106, han2017deep, LI2020100181}, Gaussian process models \cite{bartok2010gaussian, PhysRevB.90.104108, PhysRevLett.114.096405, PhysRevB.95.214302, PhysRevX.8.041048}, and linear models \cite{PhysRevB.90.024101, PhysRevB.92.054113, PhysRevMaterials.1.063801, Thompson2015316, doi-10.1137-15M1054183, PhysRevMaterials.1.043603, wood2018extending}. 
The incorporation of MLPs facilitates robust global structure searches over extensive configurational spaces that would otherwise be computationally prohibitive using DFT calculations alone. 
This approach substantially reduces the risk of failing to identify the true global minimum structures.

MLPs are expected to be highly useful for performing reliable global structure searches under high-pressure conditions. 
At the same time, the development of MLPs capable of accurately predicting a broad range of structures, including many hypothetical configurations at elevated pressures, is essential. 
Although MLPs have been developed for various systems, existing models generally require modification before being applied to global structure searches. 
Even in the case of elemental silicon, a relatively simple system for which both conventional interatomic potentials \cite{PhysRevB.31.5262, PhysRevB.33.1451, Tersoff_potential, PhysRevB.46.2727, MEAM_potential} and several MLPs \cite{PhysRevLett.114.096405, PhysRevX.8.041048, Zuo2020, LI2020100181} have been proposed, existing MLPs do not exhibit sufficient accuracy to enable global structure searches across various pressure conditions, as discussed later. 
Consequently, in most situations, it is necessary to develop MLPs with sufficient accuracy for high-pressure global structure prediction prior to conducting such searches.

The first part of this study presents an efficient and robust approach for developing MLPs applicable to crystal structure prediction under various pressure conditions. 
This approach extends an iterative procedure previously employed for global structure searches at ambient pressure using the polynomial MLP \cite{HayatoWakai202323053,PhysRevB.110.224102,seko2024polynomialmachinelearningpotential}, which demonstrated high predictive accuracy across a broad range of structures and enabled the global enumeration of local minimum structures. 
The procedure combines random structure search (RSS) with MLP development, initiated from a dataset generated under diverse pressure conditions. 
As a result, the iterative process yields an MLP with high predictive power across a wide range of structures along with a set of global and local minimum structures at various pressures.

In addition to global structure searches at zero temperature, identifying globally stable structures at finite temperatures remains a significant challenge. 
To estimate phase stability at finite temperatures based on global structure searches, self-consistent phonon approaches \cite{PhysRevB.1.572, PhysRevLett.106.165501, PhysRevB.92.054301, Thermodynamics-of-crystals, choquard1967anharmonic} can be feasibly employed within practical computational limits, provided that MLPs are developed with sufficient accuracy to support such calculations. 
These approaches incorporate both harmonic and anharmonic vibrational effects and can predict temperature-induced dynamical stability in structures that are dynamically unstable at lower temperatures. 
While self-consistent phonon calculations can be readily applied to a single and simple structure, as demonstrated in previous studies \cite{PhysRevLett.100.095901, SOUVATZIS2009888, PhysRevB.87.104111, PhysRevB.101.115119}, their application to the broad set of local minimum structures obtained from global structure searches poses several challenges: (1) MLPs must be developed with sufficient accuracy to perform self-consistent phonon calculations across a wide range of local minimum structures and thermodynamic conditions, similar to the requirements for global structure searches. 
(2) Despite the acceleration offered by MLPs, the computational cost of performing self-consistent phonon calculations across many structures and conditions remains substantial. 
(3) These calculations are demanding particularly for low-symmetry structures, which are commonly found among local minimum structures. 
Consequently, a systematic and efficient procedure for performing self-consistent phonon calculations across various structures and conditions is still needed.

The second part of this study presents a procedure for systematically and accurately performing the stochastic self-consistent harmonic approximation (SSCHA) \cite{PhysRevB.89.064302, PhysRevB.96.014111, Monacelli_2021} across a wide range of structures and thermodynamic conditions, involving the development of MLPs. 
This procedure is applied to evaluate the free energy of local minimum structures obtained from global structure searches. 
Therefore, the present study also serves as a demonstration of the applicability of SSCHA calculations to complex structures that have not been previously addressed in the literature. 
Throughout this study, it is shown that the integrated framework, comprising the generation of DFT datasets covering diverse structures and pressure-temperature conditions, the development of accurate MLPs, robust global structure searches, and reliable SSCHA calculations, can yield a highly accurate pressure-temperature phase diagram.

The performance of the proposed procedures is evaluated in elemental Si as a prototypical example, which exhibits several experimentally observed polymorphs. 
Experimental studies on elemental Si have identified seven distinct phases that exist in sequence with increasing pressure: Si-I (diamond), $\text{Si-I\hspace{-1.2pt}I}$ ($\beta$-Sn), $\text{Si-X\hspace{-1.2pt}I}$ (space group $Imma$), Si-V (simple hexagonal, SH), $\text{Si-V\hspace{-1.2pt}I}$ (space group $Cmce$), $\text{Si-V\hspace{-1.2pt}I\hspace{-1.2pt}I}$ (hexagonal close-packed, HCP), and $\text{Si-X}$ (face-centered cubic, FCC) within the pressure range of 0--100GPa at room temperature \cite{PhysRevB.50.739, PhysRevLett.82.1197,PhysRevB.41.12021}. 
Throughout this study, we have developed a novel MLP for elemental Si, starting from a DFT dataset that was previously used to construct an earlier MLP proven to be accurate for a wide variety of structures at zero pressure, including grain boundary structures \cite{FUJII2022111137} and liquid configurations \cite{PhysRevB.109.214207}. 
The current MLP also extends this accuracy to structures under pressures up to 100 GPa and significantly improves predictive performance for a broad range of local minimum structures and SSCHA calculations compared to existing MLPs.

In this study, a polynomial MLP is employed to perform the necessary calculations. 
This model is constructed using polynomial rotational invariants systematically derived from order parameters based on radial and spherical harmonic functions. Polynomial MLPs developed for general-purpose applications have demonstrated high predictive accuracy across a wide range of crystalline \cite{PhysRevB.99.214108, PhysRevB.102.174104, PhysRevMaterials.4.123607, FUJII2022111137} and liquid states \cite{PhysRevB.109.214207}. 
Although the representational capacity of the potential energy surface is constrained by the use of simple polynomial functions of the invariants, compared to more flexible models such as artificial neural networks or Gaussian process models, the model estimation remains computationally efficient. 
This efficiency is enabled by linear regression techniques supported by optimized linear algebra libraries. 
Furthermore, force and stress tensor components from DFT training datasets can be straightforwardly incorporated into the fitting procedure. 
Even when the number of entries exceeds one million due to the inclusion of these quantities, the model coefficients can still be efficiently estimated using linear regression.

\begin{figure}[!tb]
    \centering
    \includegraphics[width=0.85\linewidth]{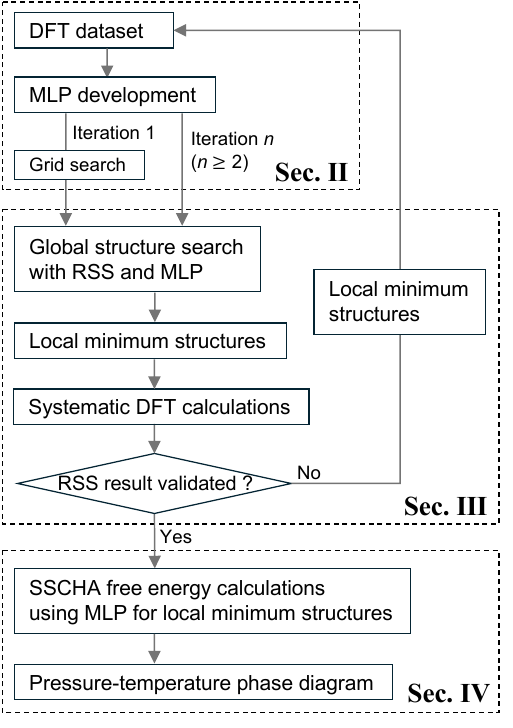}
    \caption{
    Overview of the current procedure in this study.
    The dashed boxes and corresponding section numbers indicate the sections in which the respective topics are discussed.
    }
    \label{fig:flow}
\end{figure}

An overview of the procedure proposed in this study is provided in Fig. \ref{fig:flow}, and the paper is organized as follows. 
Section \ref{MLP_development} describes the formulation and construction procedure of the polynomial MLP. 
Section \ref{MLP_optimization} presents a grid search approach for obtaining an optimal MLP, which is employed in the global structure search and phase stability analysis. 
Section \ref{Structure_enumeration} details the iterative procedure for a reliable global structure search and presents the globally stable and metastable structures identified by the search. 
Section \ref{Phase_stability} introduces the phase stability analysis procedure and provides the pressure-temperature phase diagram calculated using the proposed methodology. 
Section \ref{Performance_of_SSCHA} assesses both the computational requirements and the predictive accuracy of the MLP for SSCHA calculations. 
Finally, Sec. \ref{conclusion} concludes the study.

\section{MLP development}\label{MLP_development}
\subsection{Formulation of the polynomial MLP}
\label{MLP_discription}

In this section, the formulation of the polynomial MLP in elemental systems is presented \cite{PhysRevB.102.174104,doi:10.1063/5.0129045}.
The short-range part of the potential energy for a structure, $E$, is assumed to be decomposed as $E = \sum_i E^{(i)}$, where $E^{(i)}$ denotes the contribution of interactions between atom $i$ and its neighboring atoms within a given cutoff radius $r_c$, referred to as the atomic energy.
The atomic energy is then approximately given by a function of invariants $\{d_{m}^{(i)}\}$ with any rotations centered at the position of atom $i$ as
\begin{equation}
\label{sscha:Eqn-atomic-energy-features}
E^{(i)} = F \left( d_1^{(i)}, d_2^{(i)}, \cdots \right),
\end{equation}
where $d_{m}^{(i)}$ can be referred to as a structural feature for modeling the potential energy, and the polynomial MLP employs polynomial functions as function $F$.

When the neighboring atomic density is described by radial functions $\{f_n\}$ and spherical harmonics $\{Y_{lm}\}$, a $p$th-order polynomial invariant for radial index $n$ and set of angular numbers $\{l_1,l_2,\cdots,l_p\}$ is given by a linear combination of products of $p$ order parameters, expressed as
\begin{equation}
\label{sscha:Eqn-invariant-form}
\begin{aligned}
&d_{nl_1l_2\cdots l_p,(\sigma)}^{(i)} \\ &=
\sum_{m_1,m_2,\cdots, m_p} c^{l_1l_2\cdots l_p,(\sigma)}_{m_1m_2\cdots m_p}
a_{nl_1m_1}^{(i)} a_{nl_2m_2}^{(i)} \cdots a_{nl_pm_p}^{(i)},
\end{aligned}
\end{equation}
where the order parameter $a^{(i)}_{nlm}$ is component $nlm$ of the neighboring atomic density of atom $i$.
The coefficient set $\{c^{l_1l_2\cdots l_p,(\sigma)}_{m_1m_2\cdots m_p}\}$ ensures that the linear combinations are invariant for arbitrary rotations, which can be enumerated using group-theoretical approaches such as the projection operator method \cite{el-batanouny_wooten_2008, PhysRevB.99.214108}.
They are distinguished by index $\sigma$ if necessary.
The order parameter of atom $i$, $a_{nlm}^{(i)}$, is approximately evaluated from the neighboring atomic distribution of atom $i$ as
\begin{equation}
a_{nlm}^{(i)} = \sum_{\{j | r_{ij} \leq r_c\} }
f_n(r_{ij}) Y_{lm}^* (\theta_{ij}, \phi_{ij}),
\end{equation}
where $(r_{ij}, \theta_{ij}, \phi_{ij})$ denotes the spherical coordinates of neighboring atom $j$ centered at the position of atom $i$.
The current polynomial MLPs adopt a finite set of Gaussian-type radial functions modified by a cosine-based cutoff function to ensure smooth decay of the radial function \cite{doi:10.1063/5.0129045}.

Given a set of structural features $D^{(i)} = \{d_1^{(i)},d_2^{(i)},\cdots\}$, the polynomial function $F_\xi$ composed of all combinations of $\xi$ structural features is represented as
\begin{eqnarray}
\begin{aligned}
F_1 \left(D^{(i)}\right) &= \sum_{s} w_{s} d_{s}^{(i)} \\
F_2 \left(D^{(i)}\right) &= \sum_{\{st\}} w_{st} d_{s}^{(i)} d_{t}^{(i)} \\
F_3 \left(D^{(i)}\right) &= \sum_{\{stu\}} w_{stu} d_{s}^{(i)} d_{t}^{(i)} d_{u}^{(i)},
\end{aligned}
\end{eqnarray}
where $w$ denotes a regression coefficient.
A polynomial of the polynomial invariants $D^{(i)}$ is then described as
\begin{equation}
\label{Eqn-polynomial-model1}
E^{(i)} = F_1 \left( D^{(i)} \right) + F_2 \left( D^{(i)} \right)
+ F_3 \left( D^{(i)} \right) + \cdots.
\end{equation}
In addition to the model given by Eq. (\ref{Eqn-polynomial-model1}), simpler models composed of a linear polynomial of structural features and a polynomial of a subset of the structural features are also introduced, such as
\begin{eqnarray}
\label{Eqn-polynomial-model2}
\begin{aligned}
E^{(i)} &= F_1 \left( D^{(i)} \right)
+ F_2 \left( D_{\rm pair}^{(i)} \cup D_2^{(i)} \right) \\
E^{(i)} &= F_1 \left( D_{\rm pair}^{(i)} \cup D_2^{(i)} \cup D_3^{(i)} \right),
\end{aligned}
\end{eqnarray}
where subsets of $D^{(i)}$ are denoted by
\begin{eqnarray}
D_{\rm pair}^{(i)} = \{d_{n0}^{(i)}\}, D_2^{(i)} = \{d_{nll}^{(i)}\}, D_3^{(i)} = \{d_{nl_1l_2l_3}^{(i)}\}.
\end{eqnarray}

\subsection{Hybrid polynomial MLP models}

In this study, we introduce an extended polynomial MLP model, formulated as the sum of multiple polynomial MLPs constructed using different sets of input parameters. 
We refer to this model as a hybrid polynomial MLP model. 
Such hybrid models can be implemented in a straightforward manner, particularly within the polynomial MLP framework. 
By incorporating both single and hybrid models, the overall model flexibility is enhanced, thereby increasing the potential for improving predictive performance. 
As demonstrated later, the hybrid models achieve lower prediction errors for elemental silicon compared to the single models.

The atomic energy in the hybrid polynomial MLP is expressed as the sum of two distinct MLP models, \( E^{(i)}_{\rm{mdl1}} \) and \( E^{(i)}_{\rm{mdl2}} \), given by
\begin{equation}
E^{(i)} = E^{(i)}_{\rm{mdl1}} + E^{(i)}_{\rm{mdl2}},
\end{equation}
where
\begin{eqnarray}
\begin{aligned}
E^{(i)}_{\rm{mdl1}} &= F_1(D^{(i)}_{\rm{mdl1}} )+F_2(D^{(i)}_{\rm{mdl1}})+\cdots \\
E^{(i)}_{\rm{mdl2}} &= F_1(D^{(i)}_{\rm{mdl2}} )+F_2(D^{(i)}_{\rm{mdl2}})+\cdots.
\end{aligned}
\end{eqnarray}
Here, \( D^{(i)}_{\rm{mdl1}} \) and \( D^{(i)}_{\rm{mdl2}} \) represent sets of structural features generated using different input parameter sets, such as the cutoff radius, the number of radial functions, and the truncation of polynomial invariants. 
Specifically, these parameters include the maximum angular momentum of spherical harmonics and the polynomial order of the invariants.
These hybrid polynomial MLP models extend the framework presented in Eq. (\ref{Eqn-polynomial-model1}) and are expected to efficiently include a broader range of structural features derived from various input parameters, without requiring the consideration of intersection polynomials between the individual models.

The concept of hybrid models has been employed in several previous studies.
For instance, attempts have been made to integrate MLPs with empirical interatomic potentials, such as the embedded atom method and bond-order potentials \cite{Pun2019, PhysRevMaterials.7.043601}. 
Another example involves constructing interatomic potentials using two models with different cutoff radii, where one model captures short-range interactions and the other captures long-range interactions \cite{doi:10.1021/acs.jpcc.6b02934, PhysRevB.100.195419}.
The current hybrid polynomial MLP models follow a similar concept to the latter approach utilizing different cutoff radii.

\subsection{Datasets}
\label{dft_dataset}

We begin our dataset development by using an existing dataset generated from DFT calculations, comprising approximately 12,000 structures of elemental silicon. 
This dataset, referred to hereafter as dataset 1, was previously constructed in our earlier studies \cite{doi:10.1063/5.0129045, FUJII2022111137}. 
Dataset 1 is derived solely from prototype structures optimized at zero pressure and therefore includes only a limited number of structures with small volumes. 
As a result, the MLP trained on this dataset has relatively low predictive power for various properties under high-pressure conditions, as demonstrated in Appendix \ref{appendix_optMLP}.

The procedure used to generate dataset 1 is summarized as follows. 
First, the atomic positions and lattice constants of 86 prototype structures \cite{PhysRevB.99.214108} were optimized using DFT calculations under zero pressure. 
These prototypes consist of single elements with zero oxidation state from the Inorganic Crystal Structure Database (ICSD) \cite{Zagorac:in5024}, including metallic close-packed structures, covalent structures, layered structures, and those reported as high-pressure phases. 
Next, supercells of these optimized structures were used to introduce random lattice expansions, lattice distortions, and atomic displacements. 
DFT calculations were finally performed on these supercells to generate the dataset.

To develop MLPs with higher predictive accuracy under high-pressure conditions in this study, we construct an improved dataset by generating additional structures from prototype structures optimized at elevated pressures using the following procedure.
First, the 86 prototype structures used to construct dataset 1 are optimized using DFT calculations at 25, 50, 75, and 100 GPa.
Next, supercells of these optimized prototypes are used to create candidate structures by randomly introducing lattice expansions, distortions, and atomic displacements, following the same procedure as used for generating dataset 1.
A parameter controlling the upper bounds of the magnitude for these transformations is set to 0.3 $\rm{\AA}$.
Finally, to reduce uncertainty more efficiently than random selection, we apply the structure-selection procedure described in Ref. \cite{PhysRevB.99.214108}, extracting a total of 4,000 structures from the candidate set.
The dataset comprising these structures, combined with dataset 1, will be referred to as dataset 2. 
When the dataset is used to estimate MLPs through regression, it is randomly split into training and test sets, with 90 \% allocated to training and 10 \% to testing.

In this study, we performed DFT calculations for the additional 4,000 structures, following the computational conditions used for developing dataset 1. 
DFT calculations were carried out using the plane-wave-basis projector augmented wave (PAW) method \cite{PhysRevB.50.17953,PhysRevB.59.1758} within the Perdew--Burke--Ernzerhof (PBE) exchange-correlation functional \cite{PhysRevLett.77.3865} as implemented in the \textsc{vasp} code \cite{PhysRevB.47.558,PhysRevB.54.11169,KRESSE199615}. 
The cutoff energy was set to 400 eV, and the total energies were converged to less than 10$^{-3}$ meV per supercell. 
The allowed spacing between $k$-points was approximately 0.09 $\rm{\AA}^{-1}$. 
The configuration of the valence electrons in the PAW potential for Si is $3s^{2}3p^{2}$.

\begin{figure}[!tb]
    \centering
    \includegraphics[width=0.85\linewidth]{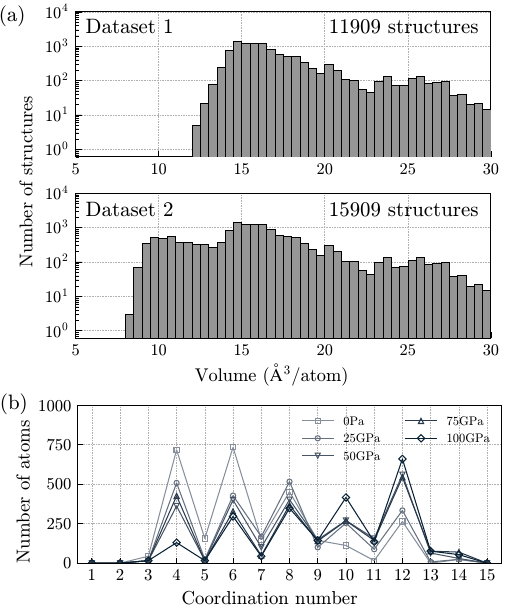}
    \caption{
    (a) Volume distributions for the structures in datasets 1 and 2, displayed on a logarithmic scale for better visibility.
    (b) Distributions of the coordination numbers around atoms in the optimized prototype structures at 0, 25, 50, 75, and 100 GPa. 
    The distributions for structures optimized under higher pressures are shown with darker lines.}
    \label{fig:dataset}
\end{figure}

Figure \ref{fig:dataset}(a) shows the volume distributions for the structures in datasets 1 and 2. 
Dataset 2 combines all structures from dataset 1 with additional small-volume structures derived from prototype structures optimized under high pressures.
Figure \ref{fig:dataset}(b) shows the distributions of the coordination numbers around the atoms in the optimized prototype structures at 0, 25, 50, 75, and 100 GPa.
We define the coordination number by counting neighbors within 1.2 times the nearest-neighbor atomic distance for each structure.
These optimized prototypes exhibit diverse neighborhood environments and coordination numbers, which tend to increase as the optimization pressure rises.

\subsection{MLP estimation}
\label{model_estimation}

The regression coefficients of a potential energy model are estimated through linear regression using energy values, force components, and stress tensor components from the training dataset as observations. 
These quantities constitute the observation vector $\bm{y}$, while the corresponding structural features and their derivatives related to the force and stress tensors form the predictor matrix $\bm{X}$:
\begin{eqnarray}
\label{MLP_component}
    \bm{X} =
    \begin{bmatrix}
    \bm{X}_{\rm{energy}} \\
    \bm{X}_{\rm{force}} \\
    \bm{X}_{\rm{stress}}
    \end{bmatrix},\quad
    \bm{y} = 
    \begin{bmatrix}
    \bm{y}_{\rm{energy}} \\
    \bm{y}_{\rm{force}} \\
    \bm{y}_{\rm{stress}}
    \end{bmatrix}.
\end{eqnarray}
Here, $\bm{X}_{\rm{energy}}$ contains polynomial features and their polynomial contributions to the total energy.
The structural features for the force $\bm{X}_{\rm{force}}$ and the stress tensor $\bm{X}_{\rm{stress}}$ are derived from derivatives of structural features, and their formulations were presented in Ref. \cite{PhysRevB.99.214108}. 
The observation vector \( \bm{y} \) contains \( \bm{y}_{\text{energy}} \), \( \bm{y}_{\text{force}} \), and \( \bm{y}_{\text{stress}} \), which consist of total energy entries, force entries, and stress tensor entries from the training dataset, respectively, computed using DFT calculations. 
The energy values and the stress tensor components are expressed in eV/cell, while the force components are expressed in eV/${\rm{\AA}}$.

Weighted linear ridge regression is employed to estimate the regression coefficients \(\bm{w}\).
This regression method incorporates a penalty term to shrink the regression coefficients while minimizing the penalized residual sum of squares, expressed as
\begin{equation}
    L({\bm{w}}) = ||{\bm{W}}({\bm{X}}{\bm{w}} - {\bm{y}}) ||^2_2 + \lambda || \bm{w} ||^2_2.
\end{equation} 
The solution for \(\bm{w}\) is obtained by solving the normal equations
\begin{equation}
(\bm{X}^T \bm{W}^2 \bm{X} + \lambda \bm{I}){\bm{w}} = \bm{X}^T \bm{W}^2 \bm{y}.
\end{equation} 
Here, \(\lambda\) represents the regularization parameter optimized to minimize prediction errors on the test set, and \(\bm{W}\) is a diagonal matrix whose entries weight the energy, force, and stress tensor data.
The weight settings for the energy and force entries are consistent with those proposed in Ref. \cite{PhysRevB.110.224102}. 
In these settings, larger weights are assigned to data entries with lower energy values and smaller forces, which is critical for accurately modeling structures near local minima. 
The weight for the stress tensor entry, $W(\sigma_{[i],\alpha\beta})$, where \(\sigma_{[i],\alpha\beta}\) denotes the stress component acting in the \(\beta\) direction on a plane normal to the \(\alpha\) direction in the $i$-th structure, is heuristically set as follows:
\begin{equation}
W(\sigma_{[i],\alpha\beta}) =
\begin{cases}
0.1\left(\dfrac{\epsilon_\sigma}{|\sigma_{[i], \alpha\beta}|}\right)^{0.7} \!\!& (|\sigma_{[i],\alpha\beta}| \geq \epsilon_\sigma) \\
0.1 & (|\sigma_{[i],\alpha\beta}| < \epsilon_\sigma)
\end{cases}.
\end{equation}
We set the threshold $\epsilon_\sigma$ to 1.0 eV/cell in this study.
To preserve high predictive accuracy of the regression model for the energy, force, and stress tensor data, the stress tensor entries are assigned relatively smaller weights than the energy entries.

\subsection{MLP model selection}\label{MLP_optimization}

As demonstrated in Ref. \cite{PhysRevB.99.214108}, the accuracy and computational efficiency of the polynomial MLP are trade-off properties. 
We therefore obtain a set of Pareto-optimal MLPs with different trade-offs between these properties through a grid search.
This search covers 523 single polynomial MLP models and 640 hybrid polynomial MLP models.
In the single polynomial MLP models, the cutoff radius ranges from 4 to 12 ${\rm{\AA}}$, and polynomial invariants are considered up to approximately 25,000.
In the hybrid polynomial MLP models, two separate models are combined. 
The first model uses smaller cutoff radii (4, 5, or 6 ${\rm{\AA}}$) and 3,000--25,000 polynomial invariants, capturing a wide range of coordination environments.
The second model adopts larger cutoff radii (6, 7, 8, 10, or 12 ${\rm{\AA}}$) and a varying number of polynomial invariants (1,000 to 25,000).
These polynomial MLPs are developed from dataset 2 using the \textsc{pypolymlp} code \cite{doi:10.1063/5.0129045, PYPOLYMLP}. 
Using the prediction errors of polynomial MLPs developed with dataset 2, we select an MLP model to be used in the subsequent structure enumeration. 
As described later, the selected polynomial MLP is retrained using an updated dataset that combines dataset 2 with local minimum structures identified through global structure searches.

\begin{figure}[!tb]
    \centering
    \includegraphics[width=\linewidth]{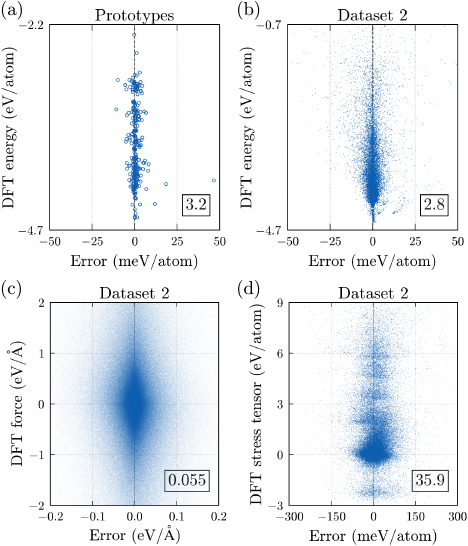}
    \caption{
    (a) Distribution of the cohesive energy for all prototype structures optimized at pressures ranging from 0 to 100 GPa.
    Distributions of (b) cohesive energy values, (c) forces, and (d) stress tensor components for dataset 2 are presented.
    They are calculated using the optimal MLP.
    The numerical values enclosed in squares represent the RMSEs for energy, force, and stress tensor components, expressed in units of meV/atom, eV/\AA, and meV/atom, respectively. 
    In panels (b), (c), and (d), the RMSEs are computed using the test dataset.
    }
    \label{fig:efs_dist}
\end{figure}

A hybrid polynomial MLP model with a computational cost of 3.1 ms/atom/step is selected from the set of Pareto-optimal MLPs, balancing prediction accuracy and efficiency. 
The root mean square errors (RMSEs) for energy, force, and stress tensor predictions are 2.8 meV/atom, 0.055 eV/\AA, and 35.9 meV/atom, respectively. 
To ensure practical relevance, these RMSE values are evaluated on test datasets derived from Dataset 2 that exclude structures with exceptionally high energies or large forces, which are assigned small weights.
Model parameters for this selected MLP are provided in the supplemental material.
The prediction errors and computational performance of other polynomial MLPs appear in Appendix \ref{appendix_MLP}. 
Although some models yield slightly lower RMSEs, their overall predictive performance on key properties remains comparable to that of the selected model. 
Furthermore, the chosen MLP model exhibits a favorable trade-off between accuracy and computational cost. 
Accordingly, this model is designated as the optimal one for the present global structure searches and phase stability analyses.

Figure \ref{fig:efs_dist}(a) compares cohesive energies for all optimized prototype structures at pressures ranging from 0 to 100 GPa calculated by both the DFT and the optimal MLP.
In most cases, the MLP yields minor errors across different pressure conditions.
Figure \ref{fig:efs_dist} also presents the distributions of (b) cohesive energy values, (c) force components, and (d) stress tensor components for dataset 2.
The MLP produces narrow error distributions in all three categories, demonstrating high predictive accuracy for diverse structures and local environments, as illustrated in Fig.~\ref{fig:dataset}(b).

The predictive performance of the optimal MLP for the energy-volume relationship, phonon properties, and thermal expansion is presented in Appendix \ref{appendix_optMLP}. 
The results demonstrate that the MLP accurately reproduces these physical properties. 
In addition, the differences in predictive accuracy between MLPs trained on dataset 1 and dataset 2 are evaluated. 
These findings indicate that achieving high predictive accuracy under elevated pressures requires the inclusion of randomly generated structures derived from prototypes that are optimized at high pressures.

\section{Structure enumeration}\label{Structure_enumeration}

\subsection{Methodology and search space}

We employ an RSS method to enumerate globally stable and metastable structures, which corresponds to the multi-start method in global optimization \cite{MultiS}.
In this approach, local geometry optimizations are repeatedly performed on random initial structures, and the local minimum structure with the lowest enthalpy is considered the global minimum.
This heuristic approach has been previously utilized in the \textit{ab initio} random structure search (AIRSS) \cite{PhysRevLett.97.045504, Ferreira2023, Pickard_2011, PhysRevLett.132.166001}.

In this study, to perform the RSS at various pressures for elemental Si, we generate initial structures by randomly assigning lattice parameters and fractional atomic coordinates in the primitive cells. 
The maximum number of atoms defining the degrees of freedom for structure optimization is restricted to 16.
We set the maximum volume per atom for an initial structure to ten times the atomic sphere volume, which is determined from the nearest-neighbor distance of the equilibrium diamond-type structure obtained from the DFT calculation.
We also discard structures that contain atomic pairs separated by excessively short distances.
This step is unavoidable because geometry optimizations starting from such structures with the current polynomial MLP often fail or converge to anomalous structures exhibiting significantly negative energy values. 
These constraints only prevent the generation of structures with excessively large volumes or extremely short interatomic distances, thereby enabling a systematic global structure search across a wide search space without relying on prior knowledge.

\subsection{Iterative update of MLP}

We utilize the polynomial MLP selected in Sec. \ref{MLP_optimization} for the global structure search. 
This MLP is selected from among Pareto-optimal MLPs to balance prediction accuracy for energy and computational efficiency.
Although the MLP exhibits high predictive power for various structures, it occasionally predicts incorrect local minimum structures and fails to accurately estimate their energy values.

To enhance the reliability of the global structure search obtained from the MLP, the MLP is iteratively updated through repeated global structure searches, as follows.
Single-point DFT calculations are performed on all local minimum structures identified by the RSS with the MLP, and these results are added to the training dataset. 
We then retrain the MLP with this expanded dataset and repeat the global structure search using the updated MLP.
This iterative approach, which combines the polynomial MLP and RSS, has been previously employed in global structure searches utilizing the polynomial MLP \cite{HayatoWakai202323053, PhysRevB.110.224102, seko2024polynomialmachinelearningpotential}.

\subsection{Computational details}

The current global structure search using the MLP is performed in three iterations.
In the first iteration, we limit the unit cell to a maximum of 8 atoms, while in the second and third iterations, we allow up to 16 atoms.
Local geometry optimizations are carried out at pressures up to 100 GPa, with a grid spacing of 5 GPa, covering a total of 21 pressure conditions.
For each pressure, we generate 1,000 initial structures for systems with up to 8 atoms and 500 initial structures for those with 9--16 atoms. 
In total, 168,000 structures are randomly generated in the first iteration, and 252,000 structures are randomly generated in the second and third iterations.
In the second and third iterations, local geometry optimizations are performed not only for newly generated random structures but also for the local minimum structures identified in previous iterations. 
This approach enables the efficient and comprehensive enumeration of local minimum structures.

\subsection{Accuracy of updated MLP}

\begin{figure*}[!tb]
    \centering
    \includegraphics[width=\linewidth]{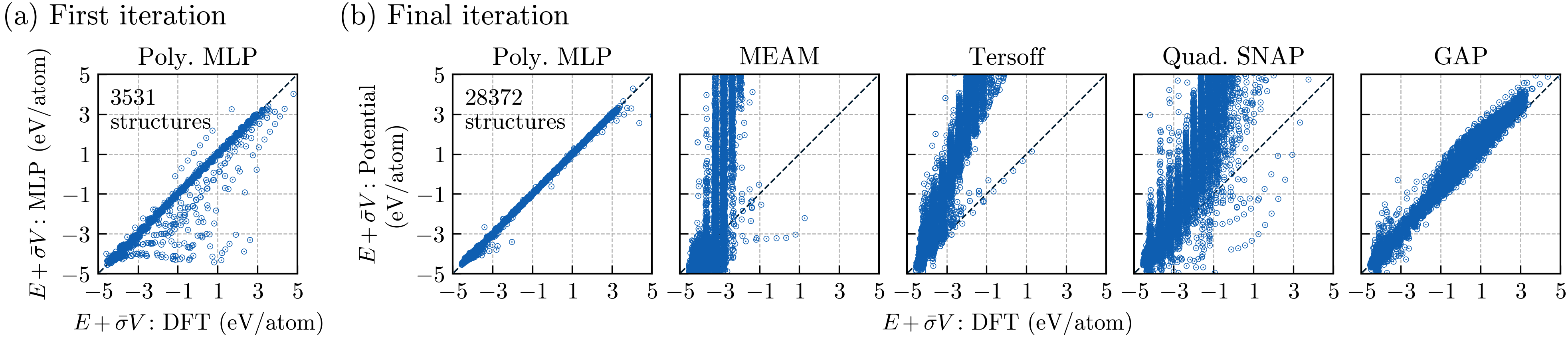}
    \caption{(a) Distributions of \( E + \bar{\sigma}V \) for local minimum structures, computed using both the DFT calculation and the polynomial MLP, at the first iteration of the iterative MLP update procedure. 
    (b) Distributions of \( E + \bar{\sigma}V \) for local minimum structures, computed using the DFT calculation and the polynomial MLP, at the final iteration. 
    For comparison, distributions computed using various interatomic potentials are also shown, including the MEAM potential \cite{MEAM_potential}, the Tersoff potential \cite{Tersoff_potential}, the quadratic SNAP \cite{Zuo2020}, and the GAP \cite{PhysRevX.8.041048}. 
    The local minimum structures used in these evaluations are obtained from the RSS with the polynomial MLP at the final iteration.
    In the figure, “Poly. MLP” and “Quad. SNAP” denote the polynomial MLP and the quadratic SNAP, respectively.}
    \label{fig:csp_iteration}
\end{figure*}

Figure \ref{fig:csp_iteration}(a) and \ref{fig:csp_iteration}(b) show the distributions of \( E + \bar{\sigma}V \) for local minimum structures obtained in the first and final iterations, evaluated by the polynomial MLPs and DFT calculations.
The distribution for each iteration is obtained using the polynomial MLP applied at the corresponding iteration.
Here, \( E \) represents the cohesive energy, \( V \) denotes the cell volume, and \( \bar{\sigma} \) is the mean normal stress, defined as \( \bar{\sigma} = (\sigma_{11} + \sigma_{22} + \sigma_{33}) / 3 \) \cite{fjaer2008petroleum}.
In the first iteration, many local minimum structures exhibit large prediction errors. 
However, by incorporating these structures and their DFT-computed properties into the training dataset and retraining the polynomial MLP, the predictive accuracy for local minimum structures improves significantly in the final iteration.

The enthalpy of a local minimum structure is given by \( E + pV \) when the stress tensor satisfies \( \sigma_{\alpha\beta} = p \delta_{\alpha\beta} \), where \( \delta_{\alpha\beta} \) is the Kronecker delta.
In Fig. \ref{fig:csp_iteration}, the stress tensor of the local minimum structure optimized by the polynomial MLP is evaluated through a single-point DFT calculation. 
Because of prediction errors in the polynomial MLP, DFT stress tensor components deviate from the condition \( \sigma_{\alpha\beta} = p \delta_{\alpha\beta} \). 
Consequently, the DFT enthalpy value of the local minimum structure cannot be directly obtained from a single-point DFT calculation.
Therefore, Fig. \ref{fig:csp_iteration} instead presents the distributions of the alternative measure \(E + \bar{\sigma}V\), which serves as an approximate evaluation of the predictive accuracy for enthalpy.

\subsection{Comparison with existing potentials}

Figure \ref{fig:csp_iteration}(b) also exhibits the distributions of \( E + \bar{\sigma}V \) for local minimum structures obtained using the polynomial MLP in the final iteration, as evaluated with various interatomic potentials.
These include the modified embedded atom method (MEAM) potential \cite{MEAM_potential}, the Tersoff potential \cite{Tersoff_potential}, the quadratic spectral neighbor analysis potential (SNAP) \cite{Zuo2020}, and the Gaussian approximation potential (GAP) \cite{PhysRevX.8.041048}.
All of these potentials fail to accurately predict the energies of a substantial number of the local minimum structures.
The mean absolute errors (MAEs) in \( E + \bar{\sigma}V \) exceed 5000 meV/atom for the MEAM potential, the Tersoff potential, and the quadratic SNAP. 
In contrast, the GAP yields a lower MAE of 267 meV/atom, while the polynomial MLP achieves a substantially reduced MAE of only 19 meV/atom. 
Moreover, a rough estimate of computational efficiency suggests that the polynomial MLP is approximately three times more efficient than the GAP. 
Such an MLP with low prediction errors for local minimum structures is essential for global structure searches.

Note that, in these comparisons, the properties obtained from our DFT calculations are used as reference values. 
While the current polynomial MLP was trained on datasets generated from our own DFT calculations, the other interatomic potentials were developed using different reference datasets based on their respective methodologies. 
The predictions of interatomic potentials derived from DFT data depend on the computational methods and settings employed, such as the choice of exchange-correlation functional. 
Moreover, when potentials are constructed based on experimental data, their predictions may deviate from those obtained within our DFT framework. 
Consequently, it is challenging to make entirely fair comparisons between the predictive performance of our polynomial MLP and that of other potentials.
Nevertheless, since our DFT calculations follow widely accepted computational settings, it is reasonable to conclude that our polynomial MLP predicts the properties of a broad range of local minimum structures with an accuracy comparable to that of conventional DFT calculations.

\subsection{Globally stable and metastable structures}

\begin{figure*}[!tb]
    \centering
    \includegraphics[width=\linewidth]{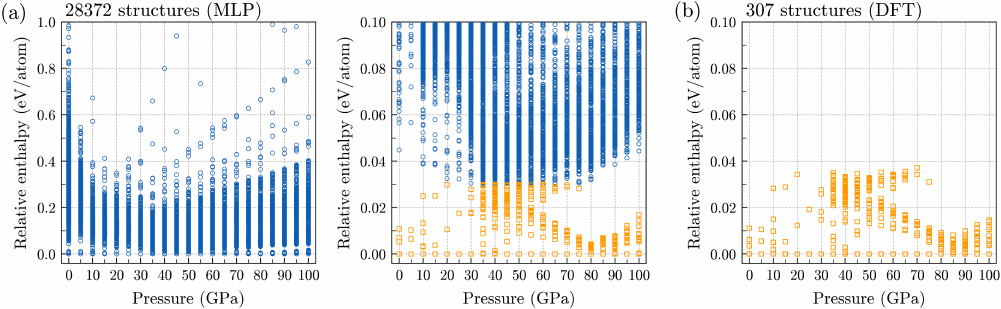}
    \caption{
    (a) Relative enthalpy values of local minimum structures computed using the MLP for elemental Si. 
    In the left and central panels, the maximum values of the vertical axis are set to 1.0 and 0.1 eV/atom, respectively. 
    The orange open squares represent structures with relative enthalpy values below 30 meV/atom, as computed by the MLP.
    (b) Relative enthalpy values of the structures indicated by the orange open squares in (a), obtained through geometry optimizations using DFT calculations. 
    }
    \label{fig:csp_mlp_Si}
\end{figure*}

By performing the RSS with the updated MLP for elemental Si, we enumerate a total of 28,372 local minimum structures.
In the final iteration, the total number of energy and force evaluations required for geometry optimizations of random structures using the MLP is 3,062,335,545. 
The MLP achieves accurate and highly efficient evaluations that would be computationally prohibitive with the DFT.
Figure \ref{fig:csp_mlp_Si}(a) presents their relative enthalpy values computed by the MLP.
At each pressure, the relative enthalpy is defined with respect to the lowest enthalpy value among the local minimum structures found at the corresponding pressure.
The MLP enables the enumeration of numerous structures across all pressure settings.
However, the enthalpy values predicted by the MLP still exhibit small but non-negligible prediction errors. 
Therefore, we compute their enthalpy values using DFT calculations to evaluate the phase stability among these local minimum structures, as follows.
First, we select 344 local minimum structures with MLP-calculated relative enthalpies below 30 meV/atom, as shown in Fig. \ref{fig:csp_mlp_Si}(a). 
We then perform DFT-based geometry optimizations for these structures and eliminate duplicates.

Geometry optimizations were performed using DFT calculations under more demanding computational conditions compared to those applied in the MLP development.
The cutoff energy was set to 600 eV, and the allowed spacing between $k$-points was approximately set to 0.07 ${\rm{\AA}}^{-1}$. 
For the local minimum structures with relative enthalpy values below 10 meV/atom at specific pressure conditions, we further relaxed them using a finer $k$-point grid spacing of 0.05 ${\rm{\AA}}^{-1}$.
The atomic positions and lattice constants were optimized until the residual forces were less than $10^{-4}$ eV/${\rm{\AA}}$.

Figure \ref{fig:csp_mlp_Si}(b) presents the DFT-calculated relative enthalpy values of 307 local minimum structures.
As shown in Figs. \ref{fig:csp_mlp_Si}(a) and (b), the MLP-calculated relative enthalpy distribution closely agrees with that computed using the DFT.
This consistency demonstrates that combining the MLP with the RSS accurately enumerates local minimum structures with low relative enthalpy values.

\begin{table}[!tb]
\centering
\caption{
\label{tab:table_Si_stable} 
Global minimum structures identified through the current global structure search for elemental Si.
The pressure ranges where these structures are global minima, as estimated from DFT calculations, are shown.
The experimentally reported pressure ranges at room temperature are also provided \cite{PhysRevB.50.739, PhysRevLett.82.1197, PhysRevB.41.12021}.
The pressure ranges are expressed in GPa.
$Z$ denotes the number of atoms in the unit cell. 
Although the $\text{Si-X\hspace{-1.2pt}I}$ structure is not found in the global structure search, the pressure range where the $\text{Si-X\hspace{-1.2pt}I}$ structure is the global minimum is evaluated and indicated in parentheses.
}
\begin{ruledtabular}
\begin{tabular}{cccccc}
 Space group & $Z$ & ICSD-ID  & Prototype & Calc. & Exp. \\ 
 \hline
 $Fd\bar{3}m$ & 8 & 51688 & Si-I (diamond) & 0--9.8 & 0--11 \\
 $I4_1/amd$ & 4 & 52460 & Si-I\hspace{-1.2pt}I ($\beta$-Sn) & 9.8--11.3 & 11--13 \\
 $Imma$ & 4 & 41392 & Si-X\hspace{-1.2pt}I & (10.6--11.4) & 13--16 \\
 $P6/mmm$ & 1 & 52456 & Si-V (SH) & 11.3--33.5 & 16--38  \\
 $Cmce$ & 16 & 89414 & Si-V\hspace{-1.2pt}I & 33.5--39.6 & 38--42  \\
 $P6_3/mmc$ & 2 & 52459 & Si-V\hspace{-1.2pt}I\hspace{-1.2pt}I (HCP) & 39.6--83.8 & 42--79  \\
 $P6_{3}/mmc$ & 4 & --- & $\alpha$-La & 83.8--87.7 & ---  \\
 $Fm\bar{3}m$ & 4 & 52458 & Si-X (FCC) & 87.7-- & 79-- \\
 \end{tabular}
\end{ruledtabular}
\end{table}

To evaluate the stability of the local minimum structures across the entire pressure range of 0--100 GPa, we compute energy values at various volumes for these structures using DFT calculations. 
We then fit these volume-energy datasets to the Rose--Vinet equation of state (EOS) \cite{PhysRevB.35.1945}, allowing us to estimate the enthalpy of each structure across the entire pressure range.
Table \ref{tab:table_Si_stable} lists the global minimum structures found within 0--100 GPa and the corresponding pressure ranges where these structures are globally stable. 
The experimentally reported pressure ranges at room temperature are also provided \cite{PhysRevB.50.739, PhysRevLett.82.1197, PhysRevB.41.12021}.
Among the structures reported experimentally, Si-I (diamond) \cite{tobbens2001}, $\text{Si-I\hspace{-1.2pt}I}$ ($\beta$-Sn) \cite{doi:10.1126/science.139.3556.762}, Si-V (SH) \cite{OLIJNYK1984137}, $\text{Si-V\hspace{-1.2pt}I}$ \cite{PhysRevLett.82.1197}, $\text{Si-V\hspace{-1.2pt}I\hspace{-1.2pt}I}$ (HCP) \cite{OLIJNYK1984137}, and Si-X (FCC) \cite{PhysRevLett.58.775} structures are calculated to be global minima, and their predicted pressure ranges closely match the experimental ones. 
The $\alpha$-La-type structure is identified as the global minimum structure, although no experimental data for this phase have been reported.

Although the $\text{Si-X\hspace{-1.2pt}I}$ structure has been reported experimentally \cite{PhysRevB.47.8337}, it was not identified in the global structure search. 
To validate this result, we performed geometry optimizations starting from the experimental $\text{Si-X\hspace{-1.2pt}I}$ structure over a fine pressure grid from 8 to 15 GPa. 
Using DFT calculations with a convergence criterion of $10^{-4}$\,eV/\AA, we found that $\text{Si-X\hspace{-1.2pt}I}$ remains a local minimum between 9.8 and 11.6\,GPa, whereas at other pressures it transforms into $\beta$-Sn- or SH-type structures. 
In contrast, under the same convergence criterion, MLP-based geometry optimizations resulted in relaxation into $\beta$-Sn- or SH-type structures over the entire 8--15 GPa range. 
This result is consistent with the absence of the $\text{Si-X\hspace{-1.2pt}I}$ structure in the MLP-based structural search. 
However, when a less strict convergence criterion of $10^{-2}$ eV/\AA\ was applied in the MLP-based optimization, the $\text{Si-X\hspace{-1.2pt}I}$ structure remained a local minimum structure within the 11.8 to 13.1 GPa range. 
This observation indicates that the enthalpy values for the $\text{Si-X\hspace{-1.2pt}I}$, $\beta$-Sn-type, and SH-type structures are closely comparable, suggesting that the potential energy surface around the $\text{Si-X\hspace{-1.2pt}I}$ structure is near that of a saddle point. 
Accurately capturing this behavior with the MLP remains a significant challenge.

To evaluate the phase stability of the $\text{Si-X\hspace{-1.2pt}I}$ structure, we obtain the enthalpy values of DFT-optimized structures in the 9.8--11.6 GPa range from the preceding validation.
Subsequently, isotropic compression is applied to the converged structure, and the corresponding energy values are computed using DFT calculations.
By fitting the volume-energy data, the EOS curve for the $\text{Si-X\hspace{-1.2pt}I}$ structure is derived. 
As a result, the $\text{Si-X\hspace{-1.2pt}I}$ structure is found to be globally stable within the 10.6--11.4 GPa range.

\begin{table}[!tb]
\centering
\caption{\label{tab:table_Si_meta}
Metastable local minimum structures, which are the same as experimentally reported structures \cite{https://doi.org/10.1002/zaac.200900051, PhysRevB.50.13043, Guerette_2020, doi:10.1126/science.193.4259.1242}. 
The relative enthalpy value ${\Delta}H$ calculated at specific pressures (in GPa) is expressed in meV/atom.
If the relative enthalpy value predicted by the MLP is less than 30 meV/atom, the relative enthalpy value calculated using DFT is provided. 
Otherwise, the value predicted by the MLP is given in parentheses.
}
\begin{ruledtabular}
\begin{tabular}{cccccc}
 Space group & $Z$ & ICSD-ID  & Prototype & ${\Delta}H$ & Pressure \\ \hline
 $Cmcm$ & 24 & 25241 & --- & (110.4) & 0 \\
 $P6_{3}/mmc$ & 4 & 30101 & Si-I\hspace{-1.2pt}V & 11.6 & 5 \\
 $Ia\bar{3}$ & 16 & 246372 & Si-I\hspace{-1.2pt}I\hspace{-1.2pt}I & (38.8) & 10 \\
 $R\bar{3}$ & 24 & 109036 & Si-X\hspace{-1.2pt}I\hspace{-1.2pt}I & (39.1) & 10 \\
 \end{tabular}
\end{ruledtabular}
\end{table}

Table \ref{tab:table_Si_meta} lists the metastable local minimum structures, which correspond to experimentally reported structures \cite{https://doi.org/10.1002/zaac.200900051, PhysRevB.50.13043, Guerette_2020, doi:10.1126/science.193.4259.1242}.
Although the $\text{Si-X\hspace{-1.2pt}I}$ structure is not identified in the current global structure search, as mentioned above, all other experimentally reported structures are discovered.
We evaluate their enthalpy values using DFT calculations if the MLP predicts enthalpies below 30 meV/atom; otherwise, we use the MLP-calculated values. 
These structures exhibit positive relative enthalpy values, which is consistent with the fact that they have not been observed as stable structures in experiments. 
The $\text{Si-I\hspace{-1.2pt}I\hspace{-1.2pt}I}$ structure has been reported as a metastable phase, obtained through slow pressure release from the high-pressure $\beta$-Sn phase \cite{https://doi.org/10.1002/zaac.200900051}. 
The $\text{Si-X\hspace{-1.2pt}I\hspace{-1.2pt}I}$ structure was also experimentally obtained by pressure release from the $\beta$-Sn phase and exists over a relatively wide pressure range of 2--12 GPa \cite{PhysRevB.50.13043}. 
The structure with space group $Cmcm$ was synthesized using a high-pressure precursor method and $\rm{Na_{4}}\rm{Si_{24}}$ precursor \cite{Guerette_2020}. 
The $\text{Si-I\hspace{-1.2pt}V}$ structure \cite{doi:10.1126/science.193.4259.1242} was synthesized by heating the $\text{Si-I\hspace{-1.2pt}I\hspace{-1.2pt}I}$ structure at 200--600 ${}^\circ$C \cite{tonkov2018phase}.
Other metastable structures with low relative enthalpy values are shown in the supplemental material, and many energetically metastable structures appear in the pressure range of 80 to 100 GPa.

\section{Phase stability at finite temperature}\label{Phase_stability}

In this section, we present a systematic procedure for performing SSCHA calculations on local minimum structures identified through global structure search. 
This procedure is applied to assess the phase stability of elemental silicon at finite temperatures. 
The SSCHA method enables a more accurate estimation of the free energy at elevated temperatures compared to the harmonic approximation. 
Furthermore, it allows for the prediction of dynamical stability in structures that appear dynamically unstable within the harmonic framework.

The SSCHA can be straightforwardly applied to a single structure with many symmetry operations, as demonstrated in the literature \cite{PhysRevB.89.064302, PhysRevB.96.184505, PhysRevB.103.134305}. 
However, the systematic application of the SSCHA to a wide range of local minimum structures remains challenging for several reasons. 
(1) A set of local minimum structures usually includes many configurations with lower symmetry. 
In such cases, SSCHA calculations require the optimization of numerous independent force constant (FC) components. 
A projector-based approach, recently developed by one of the authors, can significantly accelerate the estimation of these components \cite{PhysRevB.110.214302}, thereby enabling efficient optimization. 
(2) SSCHA calculations under various temperature and volume conditions are essential for evaluating the pressure- and temperature-dependent free energy of each local minimum structure. 
(3) Since SSCHA calculations must be performed for many local minimum structures, the total number of required computations becomes extremely large. 
(4) To efficiently manage this computational demand, it is essential to employ MLPs that accelerate energy and force evaluations during SSCHA iterations. 
To ensure sufficient accuracy in determining relative phase stability, MLPs must be developed that can accurately represent both the local minimum structures and configurations with large atomic displacements. 
The application of the SSCHA in this study establishes a robust procedure for systematic SSCHA calculations supported by MLPs.

\subsection{SSCHA formulation}

Firstly, we outline the formulation of the SSCHA for a given volume and temperature.
In the SSCHA \cite{PhysRevB.89.064302, PhysRevB.96.014111, Monacelli_2021}, the free energy is expressed by
\begin{eqnarray}
\label{SSCHA_formulation}
F_{\text{SSCHA}} = F_{\boldsymbol{\theta}} + \frac{1}{N_{\text{samp}}} \sum_{s=1}^{N_{\text{samp}}} \left[V_{\text{BO}}^{(s)} - \mathcal{V}^{(s)} \right],
\end{eqnarray}
where $F_{\boldsymbol{\theta}}$ represents the harmonic expression of the free energy, which is calculated using the effective harmonic FCs ${\boldsymbol{\theta}}$.
$V_{\text{BO}}^{(s)}$ and $\mathcal{V}^{(s)}$ correspond to the Born--Oppenheimer energy and the harmonic potential of the $s$-th sample structure, respectively.
The differences between these quantities are averaged over $N_{\rm{samp}}$ sample structures, which are generated from the density matrix at temperature $T$ with FCs ${\boldsymbol{\theta}}$.
A detailed formulation of Eq. (\ref{SSCHA_formulation}) and the density matrix can be found in Ref. \cite{PhysRevB.89.064302}. 
The SSCHA effectively captures the temperature dependence of phonon frequencies and incorporates anharmonic contributions into the free energy calculation.

We employ an iterative procedure to optimize the effective harmonic FCs \cite{VANROEKEGHEM2021107945}. 
Initially, the effective harmonic FCs are provided using a selected strategy. 
Subsequently, a displacement-force dataset consisting of $N_{\text{samp}}$ sample structures is generated using the polynomial MLP.
The updated FCs are then estimated from this dataset using standard linear regression. 
This estimation process is implemented in the \textsc{symfc} code \cite{PhysRevB.110.214302}, which facilitates efficient FC estimation even for structures with many independent FC components. 
The procedure of generating sample structures and updating the FCs is iteratively repeated until the FCs converge. 
Finally, the SSCHA free energy is obtained from Eq. (\ref{SSCHA_formulation}) using the energy and force values of the sample structures computed with the MLP.

The iterative process is terminated when the relative change in the FCs becomes smaller than a specified threshold, $\varepsilon_{\rm{FC}}$, as defined by 
\begin{eqnarray}
\label{convergence_score}
\frac{\| \boldsymbol{\theta}_{[l]} - \boldsymbol{\theta}_{[l-1]} \|_2}{\| \boldsymbol{\theta}_{[l-1]} \|_2} < \varepsilon_{\rm{FC}},
\end{eqnarray}
where $\boldsymbol{\theta}_{[l]}$ represents the FCs at iteration step $l$, and $\| \cdot \|_2$ denotes the $L_2$ norm. 
In this study, the threshold $\varepsilon_{\rm{FC}}$ is set to 0.01. 
To stabilize the convergence of the SSCHA iterations, a mixing scheme is applied when updating the FCs. 
The updated FCs are given by 
\begin{eqnarray}
\boldsymbol{\theta}_{{\rm{mix}}{[l]}} = \eta \boldsymbol{\theta}_{[l]} + (1 - \eta) \boldsymbol{\theta}_{[l-1]},
\end{eqnarray}
where the mixing parameter $\eta$ is set to 0.5 in this study.
The current procedure for SSCHA calculations, coupled with polynomial MLPs, is implemented in the \textsc{pypolymlp} code \cite{doi:10.1063/5.0129045, PYPOLYMLP}.
Calculations based on the FCs, similar to those used in the harmonic approximation, are performed using the \textsc{phonopy} code \cite{TOGO20151}.

\subsection{Computational procedures}\label{SSCHA_computational_procedure}

\begin{figure}[!tb]
    \centering
    \includegraphics[width=\linewidth]{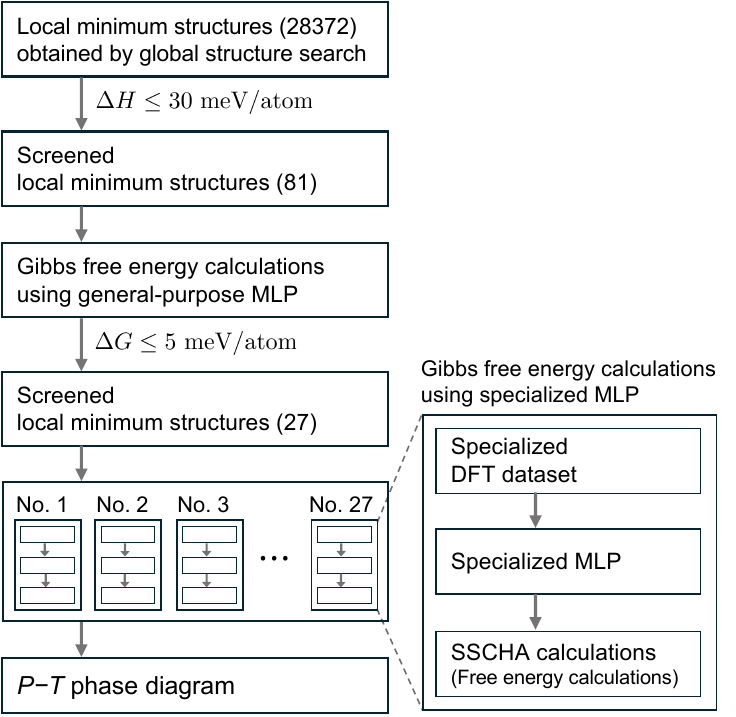}
    \caption{Workflow of the current procedure for calculating the phase diagram from the global structure search. 
    The numbers in parentheses represent the number of structures obtained in this study. 
    $\Delta{H}$ and $\Delta{G}$ denote the relative enthalpy and relative Gibbs free energy, respectively.}
    \label{fig:sscha_flow}
\end{figure}

In this study, a slightly complicated procedure is introduced to accurately evaluate phase stability while reducing computational cost. 
A workflow of the current procedure to evaluate the phase stability of local minimum structures is shown in Fig. \ref{fig:sscha_flow}. 
Firstly, 80 structures are selected from the local minima identified through the global structure search. 
These structures are those with relative enthalpy values below 30 meV/atom, as represented by the orange open squares in Fig. \ref{fig:csp_mlp_Si}(a). 
Note that the total number of orange open squares in Fig. \ref{fig:csp_mlp_Si}(a) exceeds 80 because some structures at different pressures correspond to the same structural types, such as FCC and HCP. 
After eliminating these duplicates, 80 unique structures are obtained. 
The phase stability analysis is performed on 81 local minimum structures, including the 80 unique structures in addition to the $\text{Si-X\hspace{-1.2pt}I}$ structure.

Subsequently, the volume dependence of the Helmholtz free energy is computed for the selected structures using the SSCHA over a range of temperatures. 
For each structure, 10 to 20 distinct volumes are considered, and equilibrium configurations at these volumes are determined using the MLP. 
In cases where the structure transforms into a different type during MLP-based geometry optimization, the configuration for a given volume is obtained by isotropically adjusting the volume of the structure at a reference volume where it remains stable. 
SSCHA calculations are conducted using supercells containing between 64 and 216 atoms, and the number of sample structures in each SSCHA iteration is varied between 5,000 and 20,000, depending on the number of FC components. 
The Helmholtz free energy values obtained from the systematic SSCHA calculations are then fitted to the Rose–Vinet EOS \cite{PhysRevB.35.1945}. 
The temperature range extends from 0 to 1000 K in 50 K intervals. 

As shown in Fig. \ref{fig:sscha_flow}, systematic SSCHA free energy calculations are performed twice within the workflow. 
In the first stage, we utilize the polynomial MLP selected in Sec. \ref{MLP_optimization}, which was also employed in the global structure search and demonstrates reasonable accuracy for SSCHA calculations. 
In total, Helmholtz free energies are evaluated for 23,583 unique combinations of structures, volumes, and temperatures. 
The total number of energy and force evaluations required for SSCHA calculations, performed on supercells containing between 64 and 216 atoms, amounts to 1,370,150,000. 
The use of the polynomial MLP drastically reduces the computational cost of these evaluations, thereby enabling SSCHA calculations over a broad range of local minimum structures that would be computationally prohibitive with DFT.

In the second stage, to further improve accuracy, we then develop an MLP specifically optimized for precise predictions of the target structure, following a similar approach to that used in Ref. \cite{10.1063/5.0211296} for the development of on-the-fly MLPs. 
This MLP is henceforth referred to as the specialized MLP, while the MLP used in the global structure search is referred to as the general-purpose MLP to distinguish between the two. 
These specialized MLPs are not suitable replacements for general-purpose MLPs, except when used for SSCHA calculations, because their accuracy decreases for structures other than the target one. 
However, they allow for more reliable free-energy calculations than general-purpose MLPs. 
While developing these specialized MLPs requires additional computational resources, their better performance in the large number of energy and force calculations needed for SSCHA justifies the extra computational cost.
To limit the additional computational cost associated with the DFT calculations required for training, the development of specialized MLPs is restricted to 27 structures.
These structures are selected based on their Gibbs free energy values, which are within 5 meV/atom of the lowest Gibbs free energy among the set of 81 local minimum structures. 

For each structure, we construct a DFT dataset comprising 150--300 randomly generated sample structures using the effective harmonic FCs. 
Only those FCs that have converged through SSCHA calculations with an MLP and that yield dynamically stable states are utilized.
These sample structures are selected to ensure that the specialized MLP adequately covers the entire ranges of volume and temperature. 
DFT calculations for the sample structures were carried out using the \textsc{vasp} code \cite{PhysRevB.47.558,PhysRevB.54.11169,KRESSE199615} with the same computational settings as those used for constructing the DFT dataset, as detailed in Sec. \ref{dft_dataset}.
The cutoff energy was set to 400 eV. 
The regression coefficients of the polynomial MLP were determined using weighted linear ridge regression, and leave-one-out cross-validation (CV) was utilized to estimate the prediction error, as described in Appendix \ref{appendix_CV}. 
We constructed 285 polynomial MLPs with different single polynomial MLP models. 
The specialized polynomial MLP with the lowest CV score was selected for the SSCHA calculations. 
Using the specialized MLPs, SSCHA calculations were performed over the temperature range from 0 to 1000 K.

The SSCHA calculations with the specialized MLPs require a total of 469,230,000 energy and force evaluations. 
In the entire SSCHA workflow, 1,839,380,000 supercell energy and force evaluations are performed using MLPs, whereas only 5,910 DFT energy and force evaluations are needed to develop the specialized MLPs.
This on-the-fly training strategy enables accurate SSCHA calculations while significantly reducing the computational cost of DFT calculations.

\subsection{Phase stability evaluated by the SSCHA}

\begin{figure}[!tb]
    \centering
    \includegraphics[width=0.9\linewidth]{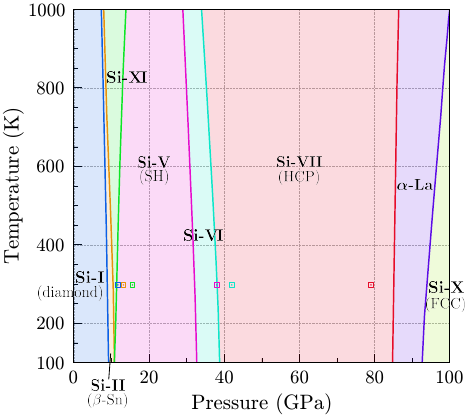}
    \caption{Pressure-temperature phase diagram calculated using SSCHA calculations with the specialized MLPs for elemental Si.
    Experimental phase transition pressures \cite{PhysRevB.50.739, PhysRevLett.82.1197,PhysRevB.41.12021} are shown as open squares, with each square colored to match the corresponding phase boundary for the phase transition observed in the experiments.
    }
    \label{fig:sscha_stable}
\end{figure}

Firstly, we perform SSCHA calculations to compute the Gibbs free energy for 81 local minimum structures with relative enthalpy values less than 30 meV/atom using the general-purpose MLP. 
Of these 81 local minimum structures, 80 are predicted to be dynamically stable within the SSCHA, which allows us to compute their free energy values. 
In contrast, only 69 structures are dynamically stable within the harmonic approximation. 
This result highlights that the SSCHA facilitates free energy calculations for structures stabilized by accounting for the temperature dependence of effective phonon frequencies.

Next, we conduct SSCHA calculations using specialized MLPs to obtain more accurate Gibbs free energy values for 27 local minimum structures. 
The Helmholtz free energies of these 27 structures calculated using the MLPs are shown in Appendix \ref{appendix_sscha_results}.
Figure \ref{fig:sscha_stable} shows the pressure-temperature phase diagram for elemental Si calculated using the SSCHA with the specialized MLPs. 
The transition pressures at room temperature calculated using the MLPs are similar to the experimental values. 
At higher temperatures, the pressure ranges where the $\text{Si-X\hspace{-1.2pt}I}$, $\text{Si-V\hspace{-1.2pt}I\hspace{-1.2pt}I}$, and $\alpha$-La-type structures are stable expand. 
The stable structures identified in the phase diagram correspond to those computed at zero temperature, as listed in Table \ref{tab:table_Si_stable}.
The number of local minimum structures with low relative Gibbs free energies is provided in the supplemental material, and this number remains nearly constant with increasing temperature. 
Many structures with low relative Gibbs free energies are found in the 80 to 100 GPa range.

\subsection{Accuracy of SSCHA calculations}\label{Performance_of_SSCHA}

\subsubsection{Predictive power}

\begin{figure*}[!tb]
    \centering
    \includegraphics[width=\linewidth]{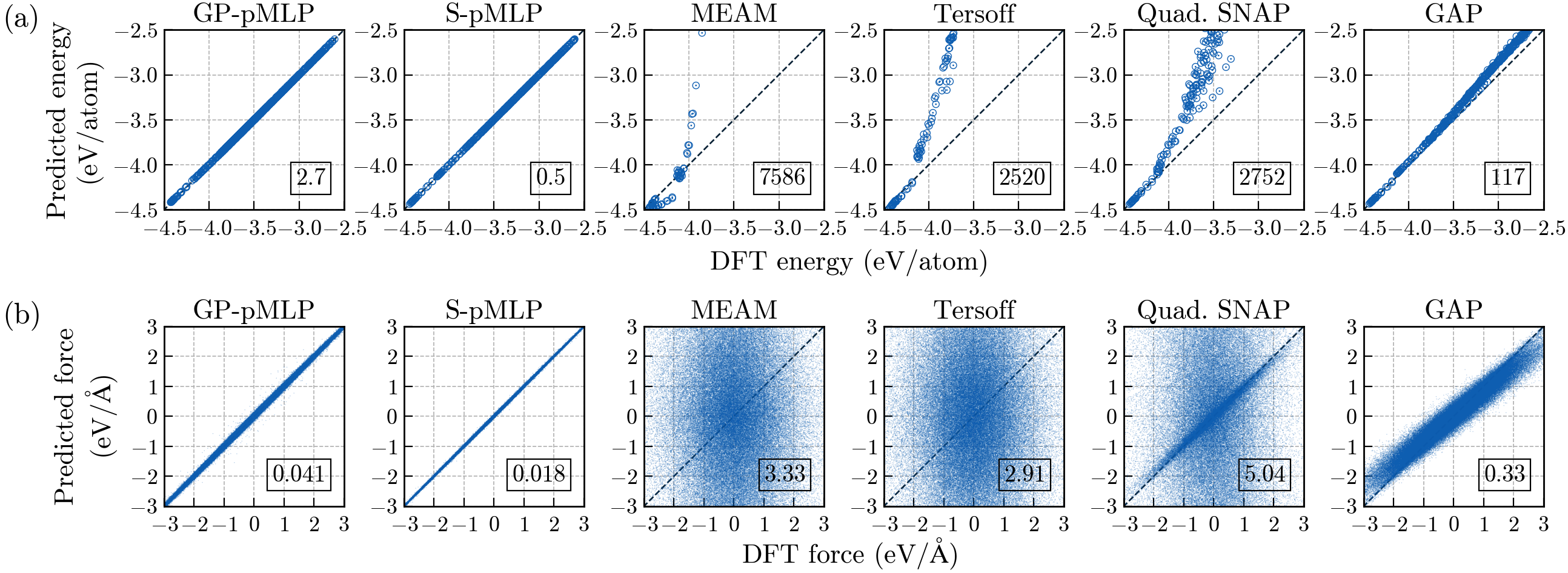}
    \caption{Distributions of (a) energies and (b) forces for sample structures in SSCHA calculations for local minimum structures. 
    The energy and force values are computed using various interatomic potentials, including the general-purpose polynomial MLP and the specialized polynomial MLPs, and are compared with those obtained from DFT calculations. 
    These potentials include the MEAM potential \cite{MEAM_potential}, the Tersoff potential \cite{Tersoff_potential}, the quadratic SNAP \cite{Zuo2020}, and the GAP \cite{PhysRevX.8.041048}. 
    The numerical values enclosed in the squares indicate RMSEs for energy and force, expressed in meV/atom and eV/$\rm{\AA}$, respectively. 
    In the figure, “GP-pMLP”, “S-pMLP”, and “Quad. SNAP” stand for general-purpose polynomial MLP, specialized polynomial MLP, and quadratic SNAP, respectively.}
    \label{fig:predict_sscha}
\end{figure*}

We evaluate the accuracy of both general-purpose and specialized MLPs in SSCHA calculations. 
In the SSCHA, energy and force values of sample structures are utilized to determine the free energy and effective FCs. 
Figure \ref{fig:predict_sscha} illustrates the distributions of (a) energies and (b) forces for representative sample structures obtained from SSCHA calculations for local minimum structures, computed using the general-purpose MLP, the specialized MLPs, and DFT calculations. 
In the panel illustrating the distribution obtained from the general-purpose MLP, one representative sample structure is selected for each combination of 81 local minimum structures and 10 to 20 volume conditions, from those generated using the converged FCs at 1000 K.
Despite many local minimum structures differing from the prototype structures used to develop the general-purpose MLP, the MLP demonstrates reasonable predictive capability for SSCHA calculations. 
This indicates that the general-purpose MLP can enable reliable SSCHA calculations across a wide range of local minimum structures. 
However, as shown in the supplemental material, prediction errors for energy values tend to be slightly larger for structures with low cohesive energies, which correspond to structures with volumes near equilibrium at low pressure. 

To evaluate the predictive accuracy of the specialized MLPs in SSCHA calculations, representative sample structures obtained from 27 local minimum structures are employed.
The predictive accuracy of the specialized MLPs is enhanced relative to that of the general-purpose MLP, thereby enabling more accurate SSCHA calculations. 
The effect of the DFT dataset size on the prediction errors in constructing the specialized MLPs for five local minimum structures is presented in Appendix \ref{appendix_dataset_effect}, demonstrating that the predictive accuracy has sufficiently converged with the current dataset size.

We also evaluate the accuracy of interatomic potentials other than the polynomial MLP in SSCHA calculations.
These include the MEAM potential \cite{MEAM_potential}, the Tersoff potential \cite{Tersoff_potential}, the quadratic SNAP \cite{Zuo2020}, and the GAP \cite{PhysRevX.8.041048}.
Figure \ref{fig:predict_sscha} also shows the distributions of (a) energies and (b) forces for the same sample structures as those used in the panel illustrating the distribution obtained from the specialized polynomial MLPs, computed using these interatomic potentials and DFT calculations.
All of these potentials fail to accurately predict the energies of the sample structures, particularly under high-pressure conditions.
Furthermore, the forces predicted by these potentials fail to reproduce those obtained from DFT calculations.

\subsubsection{Required number of sample structures}

\begin{figure}[!tb]
    \centering
    \includegraphics[width=\linewidth]{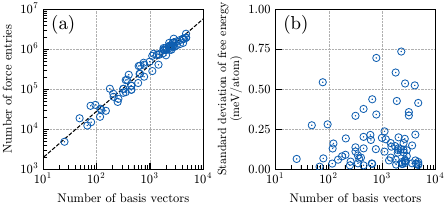}
    \caption{(a) Minimum number of force entries included in the force-displacement dataset required to achieve convergence of the FCs for the 81 local minimum structures, along with the corresponding number of basis vectors used for FC estimation. The black dotted line represents a linear fit obtained using the least-squares method on logarithmic scales.
    (b) Standard deviations of the free energy values in SSCHA calculations at 1000 K for the local minimum structures that exhibit dynamically stable states. The standard deviations are derived from free energy values obtained from 10 SSCHA trials, each initiated with random FC values.}
    \label{fig:sample_tol0.01}
\end{figure}

In the current SSCHA calculations, the effective FCs are represented by a complete orthonormal basis set \(\{\bm{b}_1, \bm{b}_2, \dots\}\), such that \(\bm{\theta} = \sum_i c_i \bm{b}_i\), where \(c_i\) denotes the expansion coefficient corresponding to the \(i\)-th basis vector.
These basis vectors satisfy the permutation symmetry rules, translational sum rules, and the symmetric properties of the supercell \cite{PhysRevB.110.214302}. 
Since a force-displacement dataset from sample structures is used to estimate the expansion coefficients during SSCHA iterations, the number of required sample structures depends on the number of basis vectors, which in turn is determined by both the number of atoms in the supercell and the symmetry properties. 
When supercells have only a limited number of symmetry operations, the number of basis vectors can become significantly large. 
Nevertheless, even for such low-symmetry structures, the basis set and the corresponding expansion coefficients can be efficiently obtained using a procedure implemented in the \textsc{symfc} code \cite{PhysRevB.110.214302}.

Figure \ref{fig:sample_tol0.01}(a) shows the minimum number of force entries in the force-displacement dataset required to achieve FC convergence for the 81 local minimum structures, along with the corresponding number of basis vectors used for FC estimation. 
These SSCHA calculations were performed using the general-purpose MLP with a convergence tolerance of 0.01. 
As depicted in Fig. \ref{fig:sample_tol0.01}(a), the number of basis vectors ranges from 25 to 4,731 across the 81 structures, and a strong correlation is observed between the number of basis vectors and the number of force entries needed for convergence. 
In these SSCHA calculations, sample structures containing 64--216 atoms were used. 
By estimating the required number of sample structures from the number of force entries, it is found that more than half of the local minimum structures require energy and force calculations for at least 1,000 sample structures per iteration to achieve FC convergence at the 0.01 tolerance. 
If more rigorous convergence tolerances are used, a larger number of sample structures is required. 
Thus, the use of polynomial MLPs facilitates SSCHA calculations for a wide variety of structures, including low-symmetry structures that require a large number of FC basis vectors.

\subsubsection{Variance of free energy}

Due to the structure sampling approach and the convergence tolerance parameter employed in the SSCHA, the calculated free energy exhibits fluctuations that depend on the specific simulation run. 
Figure \ref{fig:sample_tol0.01}(b) presents the standard deviations of the free energy values across the local minimum structures that exhibit dynamically stable states, along with the number of FC basis vectors.
The standard deviations are derived from free energy values obtained from 10 independent SSCHA trials started with random FC values. 
In each trial, the number of sample structures per iteration is set between 5,000 and 20,000, as used in the current SSCHA calculations. 
All standard deviation values are below 1 meV/atom, indicating that the current SSCHA calculations yield reasonable free energy values. 
Note that these free energy values are utilized for EOS fittings, and the fitted free energy values are subsequently used to evaluate phase stability. 
Consequently, errors arising from free energy fluctuations are expected to be minimized through the EOS fitting process.

\subsection{Comparison with previous studies}

Two previous studies by Li $\it{et~al.}$ \cite{Li_JMS_2018} and Paul $\it{et~al.}$ \cite{PhysRevLett.122.125701, PhysRevB.100.144101} reported DFT-based calculations of the pressure-temperature phase diagram of Si. 
Here, we compare our calculated pressure-temperature phase diagram with those reported in these earlier works.
Furthermore, the theoretical phase diagrams are compared with results from high-pressure dynamic-compression experiments \cite{McBride2019, PhysRevLett.121.135701, PhysRevLett.130.076101} to assess the validity of both the present and previous theoretical predictions.

\subsubsection{Comparison with theoretical phase diagrams}

The computational procedures adopted in previous phase-diagram calculations are summarized as follows.
Li $\it{et~al.}$ \cite{Li_JMS_2018} performed quasi-harmonic approximation (QHA) calculations for six structures, all of which are included in the set of stable structures predicted in the present study. 
In that study, the experimentally observed $\text{Si-X\hspace{-1.2pt}I}$ and $\text{Si-V\hspace{-1.2pt}I}$ structures were excluded.
They constructed a pressure-temperature phase diagram over the ranges of 0--2000 K and 0--80 GPa.
Paul $\it{et~al.}$ \cite{PhysRevLett.122.125701, PhysRevB.100.144101} carried out QHA calculations for ten structures, comprising seven experimentally observed phases together with the $\alpha$-La-type, body-centered cubic, and simple cubic structures. 
These experimentally observed phases and the $\alpha$-La phase correspond to the set of stable structures identified in our prediction. 
Because their QHA calculations indicated dynamical instability for the $\text{Si-X\hspace{-1.2pt}I}$ and $\text{Si-V\hspace{-1.2pt}I}$ phases, the vibrational free energies of these phases were evaluated using a molecular-dynamics-based approach \cite{VOCADLO200852}. 
A pressure-temperature phase diagram was constructed spanning temperatures up to several tens of thousands of kelvin and pressures of several terapascals.

\begin{figure*}
    \centering
    \includegraphics[width=\linewidth]{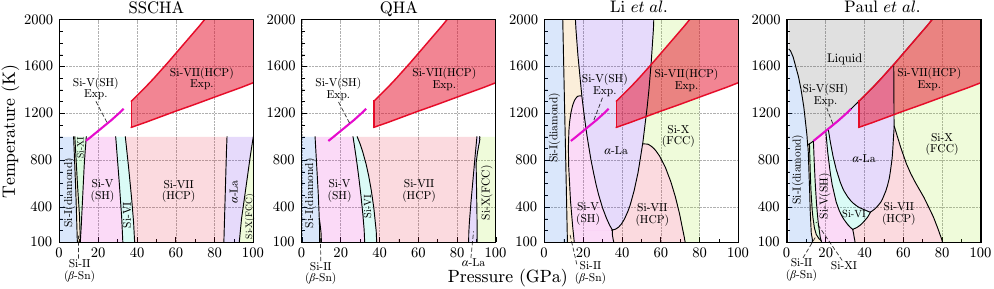}
    \caption{Pressure-temperature phase diagrams obtained from SSCHA calculations using the specialized MLPs and from QHA calculations using DFT, as well as those reported in previous studies \cite{Li_JMS_2018, PhysRevLett.122.125701, PhysRevB.100.144101}. The magenta line indicates the melting curve predicted from DFT \cite{PhysRevLett.122.125701} within the experimentally reported pressure range where the SH phase and the liquid coexist \cite{McBride2019, PhysRevLett.121.135701}.
    The region enclosed by the red lines indicates the experimentally reported stability region of the HCP phase \cite{PhysRevLett.130.076101}, which was determined with the aid of DFT calculations.}
    \label{fig:diagram_comparison_2000}
\end{figure*}

Figure \ref{fig:diagram_comparison_2000} presents the pressure-temperature phase diagram obtained from SSCHA calculations for the 80 structures enumerated in the present study.
The figure also summarizes previously reported DFT-based pressure-temperature phase diagrams \cite{Li_JMS_2018, PhysRevLett.122.125701, PhysRevB.100.144101}, reproduced from the corresponding figures in those studies for comparison.
Although the present phase diagram is consistent with earlier results below approximately 20 GPa, significant discrepancies emerge at higher pressures.
In particular, previous studies predict the $\alpha$-La-type structure to be stable over a broad range of pressures and temperatures below 60 GPa.
In contrast, in the present study, the $\alpha$-La-type structure becomes stable only within a limited pressure range of approximately 85–100 GPa at 1000 K.

To validate the accuracy of these theoretical phase diagrams, we performed DFT-based QHA calculations for the eight stable structures, employing carefully converged computational parameters.
Helmholtz free energies were obtained using the finite-displacement method implemented in the \textsc{phonopy} code \cite{TOGO20151}, with supercells containing 64--108 atoms constructed by expanding conventional unit cells of the respective structures.
DFT calculations were performed using the \textsc{vasp} code \cite{PhysRevB.47.558,PhysRevB.54.11169,KRESSE199615}.
The total energies were converged to within $10^{-5}$ meV per supercell.
To obtain the equilibrium structures, the atomic positions and lattice constants were optimized until the residual forces were less than $10^{-4}$ eV/\AA.
All other computational conditions were identical to those used for constructing the DFT dataset, as described in Sec.~\ref{dft_dataset}.

The pressure-temperature phase diagram obtained from our QHA calculations is shown in Fig.~\ref{fig:diagram_comparison_2000}. 
The phase diagrams computed using the QHA and the SSCHA exhibit only minor differences below 300 K, where anharmonic vibrational effects are expected to be small.
At higher temperatures, where anharmonic effects become more pronounced, the differences between the QHA and SSCHA phase diagrams become increasingly evident.
In contrast, the present QHA phase diagram is inconsistent with those reported in earlier studies. 
This suggests that the phase diagrams from previous studies may be less reliable, as further evidenced by the analysis of the accuracy of free energies for the stable phases.

\subsubsection{Comparison with experimental results}

\textit{In situ} X-ray diffraction studies \cite{PhysRevB.68.020102, KUBO20082255} reported stability regions above room temperature for the diamond, $\beta$-Sn, and $\text{Si-X\hspace{-1.2pt}I}$ phases. 
The phase stability behavior in these regions is consistent with that predicted by the present phase diagram.
At higher pressures, two dynamic-compression experiments by McBride $\it{et~al.}$ \cite{McBride2019} and Turneaure $\it{et~al.}$ \cite{PhysRevLett.121.135701} reported pressure ranges in which the SH-type structure coexists with the liquid phase. 
Along the melting curve, McBride $\it{et~al.}$ observed that the $\text{Si-X\hspace{-1.2pt}I}$ structure transforms into the SH-type structure at approximately 14 GPa, and that the SH phase remains stable up to approximately 27 GPa.
Turneaure $\it{et~al.}$ further reported coexistence of the SH phase and the liquid at 30.5 and 32.7 GPa.
In addition, a separate dynamic-compression experiment \cite{PhysRevLett.130.076101} proposed a stability region for the HCP phase at higher pressures, indicating that the HCP phase is stable over approximately 37–100 GPa.

The stability regions reported in the dynamic-compression experiments \cite{McBride2019, PhysRevLett.121.135701, PhysRevLett.130.076101} are shown in Fig.~\ref{fig:diagram_comparison_2000}.
We replotted these stability regions on those determined with the aid of DFT calculations \cite{PhysRevLett.122.125701, PhysRevLett.130.076101}.
Since most of the experimental results were obtained at temperatures above 1000 K, we qualitatively compare our predicted phase stability at 1000 K with the experimental observations. 
In the present SSCHA calculations, the SH phase is predicted to be stable at 1000 K over the pressure range of 13.8--29.0 GPa, which is consistent with the experimentally reported stability range. 
The SSCHA calculations also predict a broad stability field of the HCP phase at 1000 K, extending from 34.0 to 86.4 GPa.
Furthermore, the $\alpha$-La-type structure is predicted to be stable over 86.4--100.0 GPa, overlapping with the experimentally reported pressure range of the HCP phase. 
However, the free-energy difference between the $\alpha$-La-type and HCP-type structures remains small. 
Therefore, these results indicate that the present predictions are not in significant conflict with the experimental observations.

In contrast, the earlier studies predicted only a narrow stability field of the SH phase along the melting curve, in disagreement with the experimental observations.
At higher pressures, these studies predicted the stability of the $\alpha$-La-type structure, for which no experimental evidence has been reported.
Furthermore, they predicted the $\alpha$-La- or FCC-type structures to be stable within the experimentally reported stability field of the HCP phase, which is inconsistent with the experimental findings.

\begin{figure*}[!tb]
    \centering
    \includegraphics[width=0.93\linewidth]{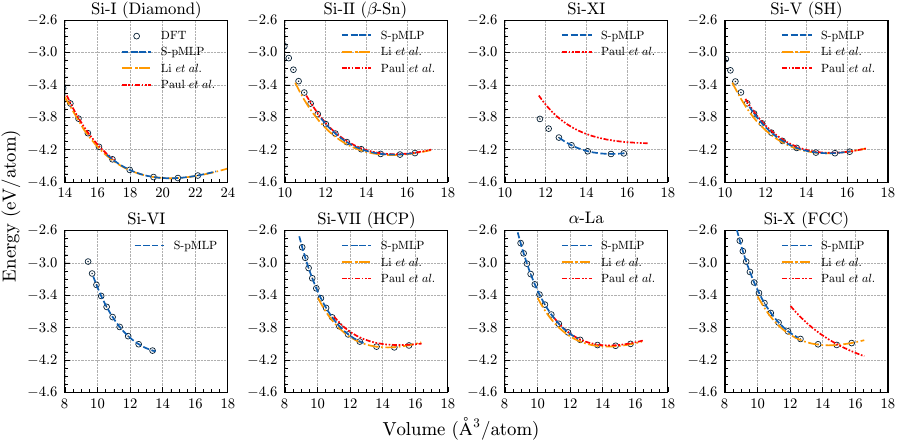}
    \caption{Energy-volume curves computed using the specialized MLPs and reported in previous studies \cite{Li_JMS_2018, PhysRevB.100.144101}.
    The curves are shown for eight globally stable structures identified in the present study, as shown in Fig.~\ref{fig:sscha_stable}. The energies obtained from our DFT calculations are shown as black points. In the legend, “S-pMLP” stands for specialized polynomial MLP.}
    \label{fig:eos_comparison}
\end{figure*}

\begin{figure*}[!tb]
    \centering
    \includegraphics[width=\linewidth]{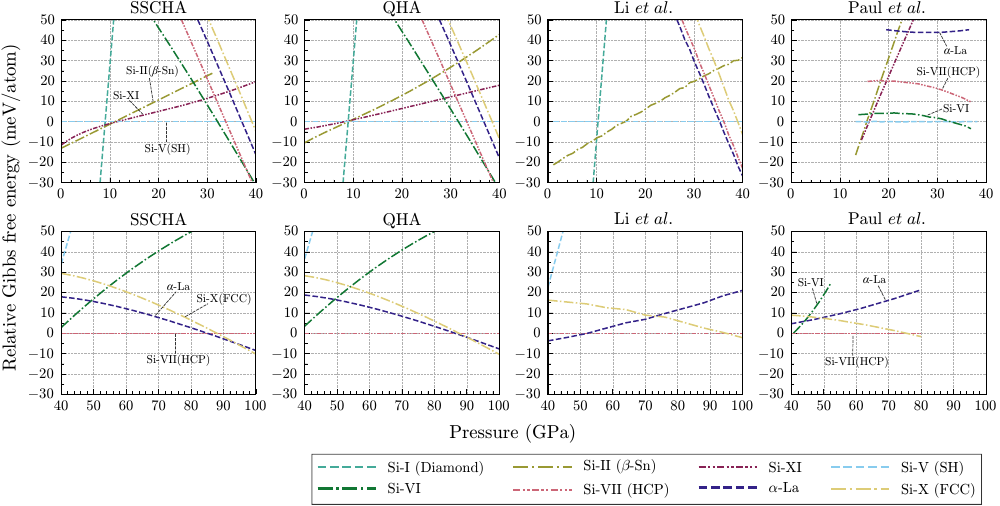}
    \caption{Relative Gibbs free energies at 300 K for the pressure ranges of 0--40 and 40--100 GPa, referenced to the SH- and HCP-type structures, respectively. This figure includes results obtained from SSCHA calculations using the specialized MLPs and from QHA calculations using DFT, as well as those reported in previous studies \cite{Li_JMS_2018, PhysRevLett.122.125701}.}
    \label{fig:gibbs_comparison}
\end{figure*}

\subsubsection{Comparison of energy-volume curves}

To examine the origin of the discrepancies among the theoretical phase diagrams, we compare the energy-volume curves and Gibbs free energies obtained in the present work with those reported in previous studies. 
Figure \ref{fig:eos_comparison} shows the energy-volume curves computed using our DFT calculations and the specialized MLPs. 
The MLPs accurately reproduce the energy values obtained from the DFT calculations. 
The figure also includes the energy-volume curves reported in previous studies \cite{Li_JMS_2018, PhysRevB.100.144101}.
Note that, because different reference energies were used in the previous studies, we adopt the energy of the equilibrium diamond-type structure at 0 GPa as a common reference in order to compare the energy-volume curves among the different works.

As shown in Fig.~\ref{fig:eos_comparison}, the energy-volume curves reported by Li $\it{et~al.}$ \cite{Li_JMS_2018} for many structures are generally consistent with the DFT energies calculated in the present study, although some deviations are observed at low-volume regions. 
These deviations may be attributed to insufficient $k$-point convergence in their calculations. 
The supplemental material presents the energies of several structures computed using different $k$-point settings, including those employed by Li $\it{et~al.}$ and those used in the present study.
While the $k$-point settings adopted in the present work yield fully converged energies, those used by Li $\it{et~al.}$ do not achieve convergence for the SH-, HCP-, $\alpha$-La-, and FCC-type structures.

Compared with the results of Paul $\it{et~al.}$ \cite{PhysRevB.100.144101}, the energy-volume curves for the diamond-, $\beta$-Sn-, SH-, and $\alpha$-La-type structures are consistent with our DFT calculations. 
In contrast, noticeable differences are observed for the HCP-type structure, and much larger discrepancies are found for the $\text{Si-X\hspace{-1.2pt}I}$ and FCC-type structures.
Since the energy-volume curve for the $\text{Si-V\hspace{-1.2pt}I}$ structure is not provided in Ref.~\cite{PhysRevB.100.144101}, it is not included in Fig.~\ref{fig:eos_comparison}.

\subsubsection{Comparison of relative Gibbs free energies}

Figure \ref{fig:gibbs_comparison} shows the relative Gibbs free energies at 300 K obtained from SSCHA calculations using specialized MLPs and those obtained from DFT-based QHA calculations.
We present these values for the pressure ranges of 0--40 and 40--100 GPa, referenced to the SH- and HCP-type structures, respectively.
The contribution from anharmonic effects should be small at 300 K, and the relative Gibbs free energies obtained from QHA and SSCHA calculations are nearly identical.

Figure \ref{fig:gibbs_comparison} also shows the relative Gibbs free energies at 300 K reported in previous studies \cite{Li_JMS_2018, PhysRevLett.122.125701}.
For most structures, the relative Gibbs free energies reported by Li $\it{et~al.}$ \cite{Li_JMS_2018} are in close agreement with those obtained in the present study. 
However, the relative Gibbs free energy of the $\alpha$-La-type structure deviates from the values obtained in this study. 
This discrepancy is likely due to insufficient $k$-point sampling in the previous study.

The relative Gibbs free energies reported by Paul $\it{et~al.}$ \cite{PhysRevLett.122.125701} differ substantially from our results in the pressure range of 0--40 GPa.
In the range of 40--100 GPa, the Gibbs free energies of the $\alpha$-La- and FCC-type structures reported by Paul $\it{et~al.}$ also deviate from those obtained in the present study.
There are two notable differences in the computational procedures between the present study and the earlier work. 
First, Paul $\it{et~al.}$ included the contribution from electronic excitations to the free energy. 
Second, they employed a PAW potential treating $2s^{2}2p^{6}3s^{2}3p^{2}$ as valence electrons. 
To assess the impact of these choices on the free energies in the high-pressure region, we performed additional calculations incorporating these contributions.
The supplemental material presents the relative Gibbs free energies of the $\alpha$-La- and FCC-type structures with respect to the HCP-type structure using the computational procedure adopted in the previous study.
Our results indicate that the contribution from electronic excitations is negligible.
The effect of using the different PAW potential is also small for both the $\alpha$-La- and FCC-type structures.
These findings suggest that the large discrepancies observed in the relative Gibbs free energies are unlikely to originate from differences in these computational settings.

In addition, the relative Gibbs free energy values reported by Paul $\it{et~al.}$ \cite{PhysRevLett.122.125701} appear to be unreliable, as inconsistencies are observed among their reported values for the $\text{Si-V\hspace{-1.2pt}I}$, HCP-type, and $\alpha$-La-type structures.
In their work, the free energies are provided for both the pressure range of 0--40 GPa and 40--80 GPa.
However, the values around 40 GPa are inconsistent with each other, as shown in Fig.~\ref{fig:gibbs_comparison}.

\begin{figure}[!tb]
    \centering
    \includegraphics[width=0.95\linewidth]{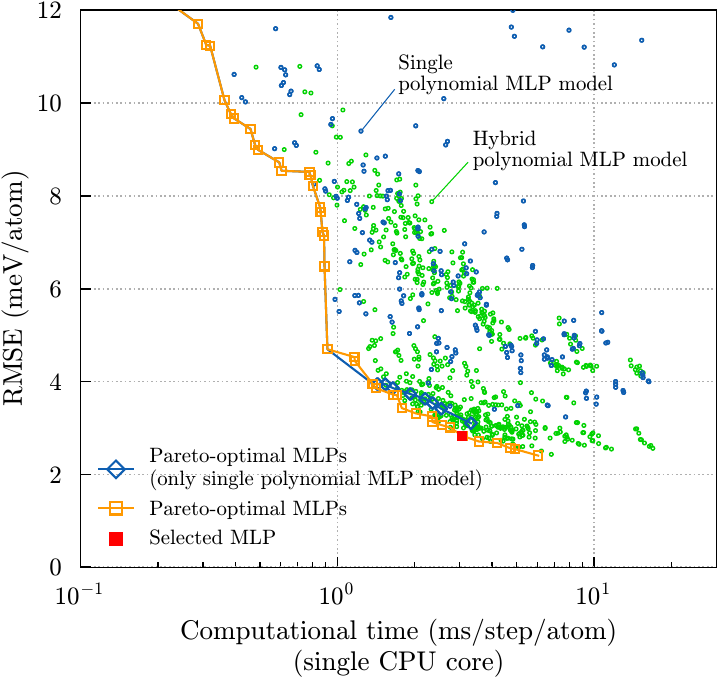}
    \caption{
    Prediction error and computational efficiency of polynomial MLPs obtained from the grid search in elemental Si.
    The blue and green open circles represent the properties of MLPs constructed using single polynomial MLP models and hybrid polynomial MLP models, respectively. 
    Computational efficiency is assessed based on the computation time using a single core of the Intel$^\text{\textregistered}$ Xeon$^\text{\textregistered}$ E5-2630 v3 (2.40 GHz). 
    The orange open squares denote the Pareto-optimal MLPs across both single and hybrid polynomial MLP models, while the blue open diamonds represent the Pareto-optimal MLPs from the single polynomial MLP models only.
    The red filled square indicates the selected MLP for conducting the global structure search and phase stability analysis.
    }
    \label{fig:rmse_time_Si}
\end{figure}

\section{Conclusion}\label{conclusion}

This study proposes a procedure for accurately assessing phase stability through global structure searches under various pressure and temperature conditions. 
The approach combines global structure searches involving numerous local geometry optimizations with systematic SSCHA calculations performed on many structures across different conditions. 
Central to this procedure is the development of reliable polynomial MLPs capable of enumerating diverse local minimum structures over a wide pressure range and enabling SSCHA calculations under combined pressure and temperature conditions. 
The developed MLPs significantly accelerate the extensive energy and force evaluations required, while maintaining high accuracy.

The performance of the proposed procedure has been demonstrated for elemental silicon as a prototypical example. 
A polynomial MLP was developed using a dataset comprising randomly generated structures derived from optimized prototype structures at pressures ranging from 0 to 100 GPa, in addition to local minimum structures obtained from RSS. 
The MLP successfully identified a wide variety of globally stable and metastable structures, including nearly all experimentally reported phases, with accuracy comparable to that of DFT calculations. 
Furthermore, systematic free-energy evaluations of the local minimum structures, incorporating anharmonic contributions, yielded a reliable pressure-temperature phase diagram spanning 0--1000 K and 0--100 GPa. 
Consequently, this study establishes a general and robust framework for reliable crystal structure prediction under pressure and finite-temperature conditions, applicable to a broad range of materials systems.

\begin{acknowledgments}
This work was supported by a Grant-in-Aid for Scientific Research (A) (Grant Number 21H04621) and a Grant-in-Aid for Scientific Research (B) (Grant Number 22H01756) from the Japan Society for the Promotion of Science (JSPS).
\end{acknowledgments}

\appendix

\section{Pareto-optimal MLPs}\label{appendix_MLP}

\begin{figure}[!tb]
    \centering
    \includegraphics[width=\linewidth]{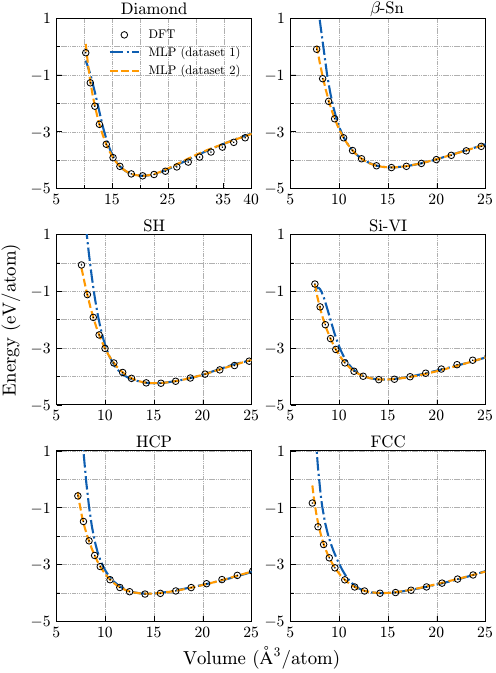}
    \caption{Energy-volume curves of the diamond-, $\beta$-Sn-, SH-, $\text{Si-V\hspace{-1.2pt}I-}$, HCP-, and FCC-type structures, computed using the DFT calculation and MLPs developed from datasets 1 and 2 for elemental Si. 
    Dataset 2 includes random structures derived from prototype structures optimized at high pressures.}
    \label{fig:EV_plot}
\end{figure}

\begin{figure*}[!tb]
    \centering
    \includegraphics[width=0.95\linewidth]{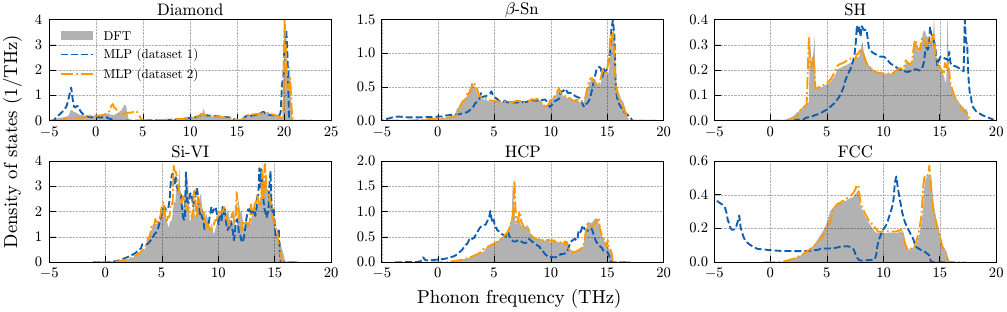}
    \caption{Phonon DOS for diamond-, $\beta$-Sn-, SH-, $\text{Si-V\hspace{-1.2pt}I-}$, HCP-, and FCC-type structures at 50 GPa in elemental Si. 
    These are computed using DFT calculations and MLPs developed using datasets 1 and 2 within the harmonic approximation. 
    The shaded region represents the phonon DOS obtained using the DFT calculation.}
    \label{fig:all_dos}
\end{figure*}

\begin{figure}[!tb]
    \centering
    \includegraphics[width=\linewidth]{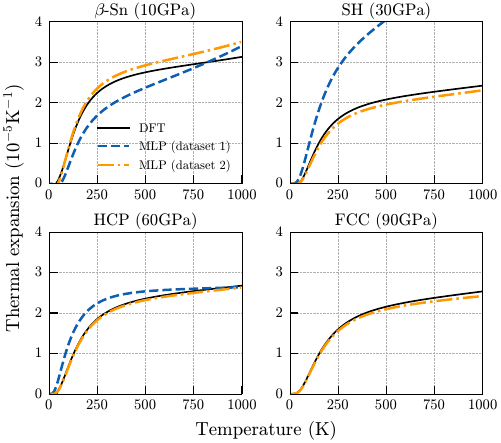}
    \caption{Temperature dependence of the thermal expansion in $\beta$-Sn-, SH-, HCP-, and FCC-type structures at 10, 30, 60, and 90 GPa, respectively, for elemental Si.
    These are computed using DFT calculations and MLPs developed using datasets 1 and 2 within the QHA. 
    }
    \label{fig:sum_therm_exp}
\end{figure}

Figure \ref{fig:rmse_time_Si} shows the prediction error and computational efficiency of the polynomial MLPs obtained through a grid search of 523 single-polynomial and 640 hybrid-polynomial MLP models for elemental Si.
These MLPs are constructed using dataset 2, which includes approximately 16,000 structures shown in Fig. \ref{fig:dataset}.
The prediction errors are evaluated using the RMSEs of the energy.
The computational efficiency is assessed by measuring the elapsed time to compute the energy, forces, and stress tensors ten times for a structure containing 1,000 atoms. 
As the model complexity of the polynomial MLP increases, the prediction error generally decreases.
Among the Pareto-optimal MLPs, the polynomial MLPs constructed using hybrid polynomial MLP models demonstrate lower prediction errors than those constructed with single polynomial MLP models when the computational cost exceeds 1.5 ms/step/atom. 
Hence, the hybrid polynomial MLPs can provide Pareto-optimal MLPs with a better trade-off between prediction error and computational cost.
The optimal MLP selected for global structure searches and phase stability analyses is indicated by the red filled square in Fig. \ref{fig:rmse_time_Si}.

\section{Predictive power of polynomial MLPs}\label{appendix_optMLP}

\begin{figure}[!tb]
    \centering
    \includegraphics[width=0.95\linewidth]{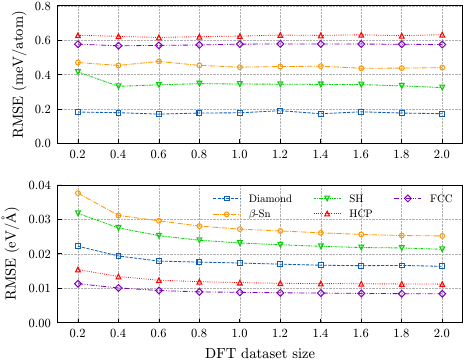}
    \caption{Relationship between the RMSE evaluated by the cross-validation approach and the DFT dataset size for five local minimum structures. The dataset sizes are expressed as multiples of the current DFT dataset size used in the development of the specialized MLPs. The top and bottom panels show the prediction errors for energy and force, respectively.}
    \label{fig:cv_converge}
\end{figure}

\begin{figure*}[!tb]
    \centering
    \includegraphics[width=\linewidth]{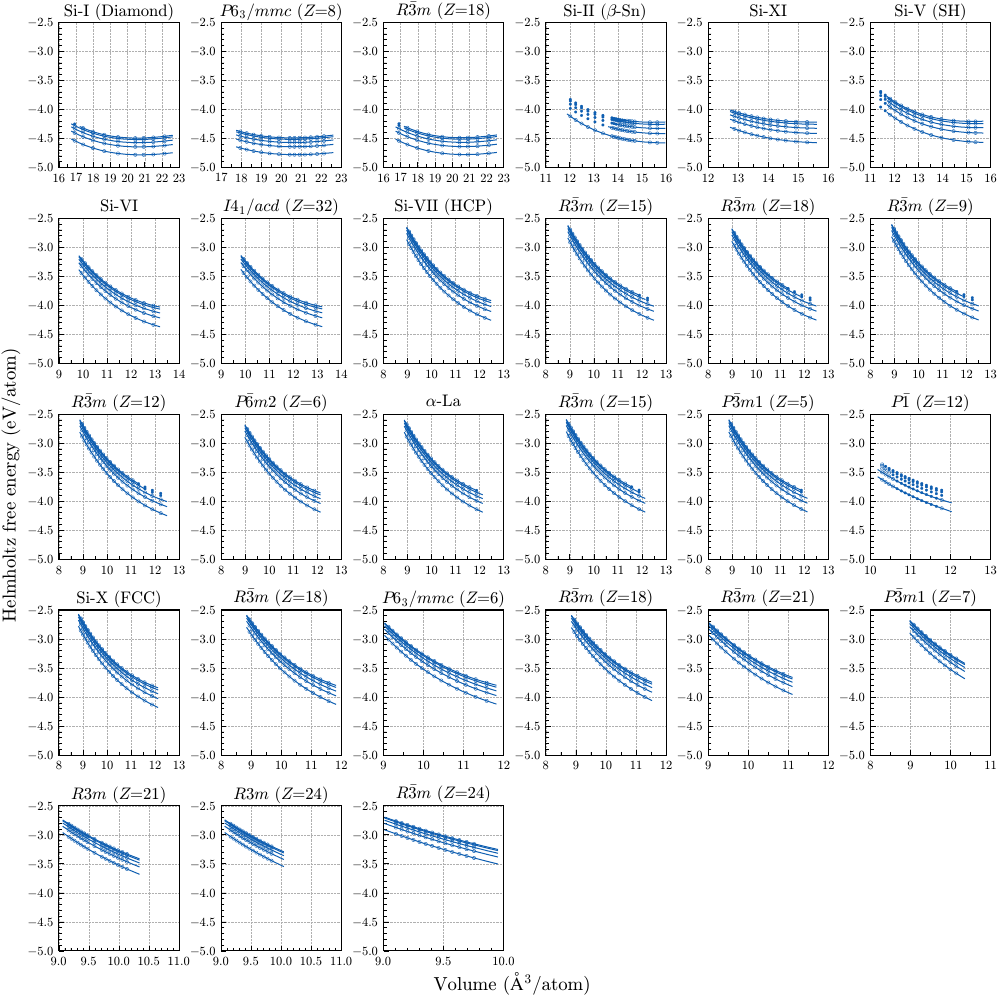}
    \caption{Helmholtz free energies for 27 structures calculated using the specialized MLPs. Dynamically stable structures are shown as open points, whereas dynamically unstable structures are shown as closed points. Although SSCHA calculations were performed over the temperature range from 0 to 1000 K with a 50 K interval, Helmholtz free energies computed at 100, 300, 500, 700, and 1000 K are shown as representative examples for clarity.}
    \label{fig:helmholtz_spmlp_sscha}
\end{figure*}

Here, we present the predictive capability of the optimal MLP for the energy-volume curve and phonon properties.
Datasets 1 and 2, shown in Fig. \ref{fig:dataset}, are used to train two different MLPs, which are then applied to calculate the energy-volume curve and phonon properties.
The key difference between these datasets is that dataset 2 includes random structures derived from prototype structures optimized under high pressure.
This section demonstrates that achieving high predictive accuracy under elevated pressures requires incorporating random structures derived from high-pressure optimized prototypes.

Figure \ref{fig:EV_plot} shows the energy-volume curves of the diamond-, $\beta$-Sn-, SH-, $\text{Si-V\hspace{-1.2pt}I-}$, HCP-, and FCC-type structures in elemental Si. 
These structures have been experimentally observed at 0--100 GPa and room temperature.
We compare energy-volume curves computed using the DFT with those from two MLPs.
In the small-volume region, the MLP constructed from dataset 1 deviates from the DFT values due to the insufficient number of small-volume structures in dataset 1.
In contrast, the MLP developed using dataset 2 accurately reproduces the DFT results for all structures.

Figure \ref{fig:all_dos} presents the phonon density of states (DOS) for diamond-, $\beta$-Sn-, SH-, $\text{Si-V\hspace{-1.2pt}I-}$, HCP-, and FCC-type structures at 50 GPa in elemental Si.
These calculations are performed using the DFT and the two MLPs within the harmonic approximation.
The finite displacement method in the \textsc{phonopy} code \cite{TOGO20151} is employed to obtain phonon properties and thermal expansion, with supercells constructed by expanding the conventional unit cells of the respective structures.
The phonon DOS calculated with the MLP constructed from dataset 1 is inconsistent with the DFT results, especially for the FCC-type structure.
By comparison, the phonon DOS obtained using the MLP from dataset 2 agrees with the DFT results.

Figure \ref{fig:sum_therm_exp} shows the temperature dependence of thermal expansion, calculated using the QHA, for the $\beta$-Sn-, SH-, HCP-, and FCC-type structures at 10, 30, 60, and 90 GPa, respectively.
Although developing MLPs that accurately predict thermal expansion is more challenging than those designed to predict phonon DOS, the MLP developed from dataset 2 accurately predicts thermal expansion.
By contrast, the MLP trained on dataset 1 produces inconsistent results and cannot compute the thermal expansion for the FCC-type structure because it predicts the FCC-type structure to be dynamically unstable.

Note that dataset 2 consists of approximately 16,000 structures, of which 4,000 are random structures derived from prototype structures optimized at high pressures. 
Even though these random structures account for only about one-fourth of the dataset, they significantly enhance the predictive power of the polynomial MLP for the energy-volume curve and phonon properties under high-pressure conditions. 
Furthermore, this improvement does not depend on any specific MLP choice, since other Pareto-optimal MLPs also show consistently high accuracy.
Therefore, the current dataset provides sufficient information to enable accurate predictions under high-pressure conditions.

\section{Leave-one-out cross validation}\label{appendix_CV}

We introduce the weighted leave-one-out CV method to estimate the prediction error of the polynomial MLP. 
The weighted leave-one-out CV score is defined by 
\begin{eqnarray}
\label{LOOCV_1}
(\text{CV score}) = \frac{1}{N_{\rm{data}}} \sum_{i=1}^{N_{\rm{data}}} \left\{w_i(  \bm{x}_i^T\hat{\bm{w}}_{(-i)}  - y_i) \right\}^2,
\end{eqnarray}
where \( N_{\text{data}} \) represents the number of the training data points. 
\( \bm{x}_i \), \( y_i \), and \( w_i \) correspond to the predictor vector, the observation value, and the weight for the \( i \)-th data point, respectively. 
The vector \( \hat{\bm{w}}_{(-i)} \) denotes the regression coefficients estimated using the dataset excluding the \( i \)-th data point.

When calculating the CV score using Eq. (\ref{LOOCV_1}), the regression coefficients must be estimated \( N_{\text{data}} \) times, which can be computationally demanding. 
However, the CV score for weighted linear ridge regression can be determined by estimating the regression coefficients only once. 
In this context, the CV score can be rewritten as 
\begin{eqnarray}
\text{(CV score)} = \frac{1}{N_{\rm{data}}} \sum_{i=1}^{N_{\rm{data}}} \left\{ \frac{w_{i}(\bm{x}_i^T \bm{w}  - y_i) }{1 - h_{ii}(\lambda)} \right\}^2,
\end{eqnarray}
where \( h_{ii}(\lambda) \) represents the \( i \)-th diagonal element of the hat matrix \( H(\lambda) \). 
The hat matrix is defined as
\begin{eqnarray} 
\bm{H}(\lambda) = \bm{X} (\bm{X}^T \bm{W}^2 \bm{X} + \lambda \bm{I})^{-1} \bm{X}^T \bm{W}^2,
\end{eqnarray} 
with the symbols explained in Sec. \ref{model_estimation}. 
Consequently, the diagonal elements of \( H(\lambda) \) can be computed sequentially using 
\begin{eqnarray}
h_{ii}(\lambda) = \bm{x}_i^T (\bm{X}^T \bm{W}^2 \bm{X} + \lambda \bm{I})^{-1} \bm{x}_i w_{i}^2.
\end{eqnarray}

\section{Effect of DFT dataset size on specialized MLPs accuracy}\label{appendix_dataset_effect}

The effect of DFT dataset size on the prediction errors in constructing the specialized MLPs is investigated.
Figure \ref{fig:cv_converge} shows the relationship between the RMSE evaluated by the cross-validation approach and the DFT dataset size for five local minimum structures. 
The dataset sizes are expressed as multiples of the current DFT dataset size used to develop the specialized MLPs, ranging from 0.2 to 2.0. 
The predictive accuracy has sufficiently converged with the current dataset size. 
Additionally, MLPs developed with dataset sizes as small as 0.4 exhibit reasonable predictive performance, indicating that even smaller datasets can yield sufficiently accurate predictions.

\section{Helmholtz free energies calculated using the specialized MLPs}\label{appendix_sscha_results}

Figure \ref{fig:helmholtz_spmlp_sscha} shows the Helmholtz free energies of 27 structures calculated using the specialized MLPs at 100, 300, 500, 700, and 1000 K. 
These results demonstrate that the MLPs enable free-energy evaluations using the SSCHA for diverse structures, including low-symmetry ones.

\bibliography{introduction, references}%

\end{document}


\preprint{APS/123-QED}

\title{Supplemental Material\\Global structure searches under varying temperatures and pressures using polynomial machine learning potentials: A case study on silicon}

\author{Hayato \surname{Wakai}}
\email{wakai@cms.mtl.kyoto-u.ac.jp}
\affiliation{Department of Materials Science and Engineering, Kyoto University, Kyoto 606-8501, Japan}
\author{Atsuto \surname{Seko}}
\email{seko@cms.mtl.kyoto-u.ac.jp}
\affiliation{Department of Materials Science and Engineering, Kyoto University, Kyoto 606-8501, Japan}
\author{Isao \surname{Tanaka}}
\affiliation{Department of Materials Science and Engineering, Kyoto University, Kyoto 606-8501, Japan}
\date{\today}

\date{\today}

\maketitle

\subsection{Model parameters of the selected polynomial MLP}

We select a hybrid polynomial MLP with a computational cost of 3.1 ms/atom/step from the set of Pareto-optimal MLPs, balancing prediction accuracy and efficiency, as discussed in Section $\text{I\hspace{-1.2pt}I\hspace{1pt}E}$ of the main text.
The root mean square errors (RMSEs) for energy, force, and stress tensor predictions are 2.8 meV/atom, 0.055 eV/\AA, and 35.9 meV/atom, respectively.
This hybrid polynomial model consists of two sub-models. 
One uses a cutoff radius of 5.0~${\rm{\AA}}$, and the other uses 6.0~${\rm{\AA}}$.
Both use seven radial basis functions and spherical harmonics truncated at $l^{(2)}_{\rm{max}} = 8$ and $l^{(3)}_{\rm{max}} = 4$.
The potential energy in each case is given by
\begin{equation}
\label{Eqn-polynomial-model1}
E^{(i)} = F_1 \left( D^{(i)} \right) + F_2 \left( D^{(i)} \right),
\end{equation}
where the total number of expansion coefficients is 18,720.
This model is designated as optimal for the present global structure searches and phase stability analyses.

\subsection{Metastable structures}\label{appendix_metastable}

\begin{figure*}
    \centering
    \includegraphics[width=\linewidth]{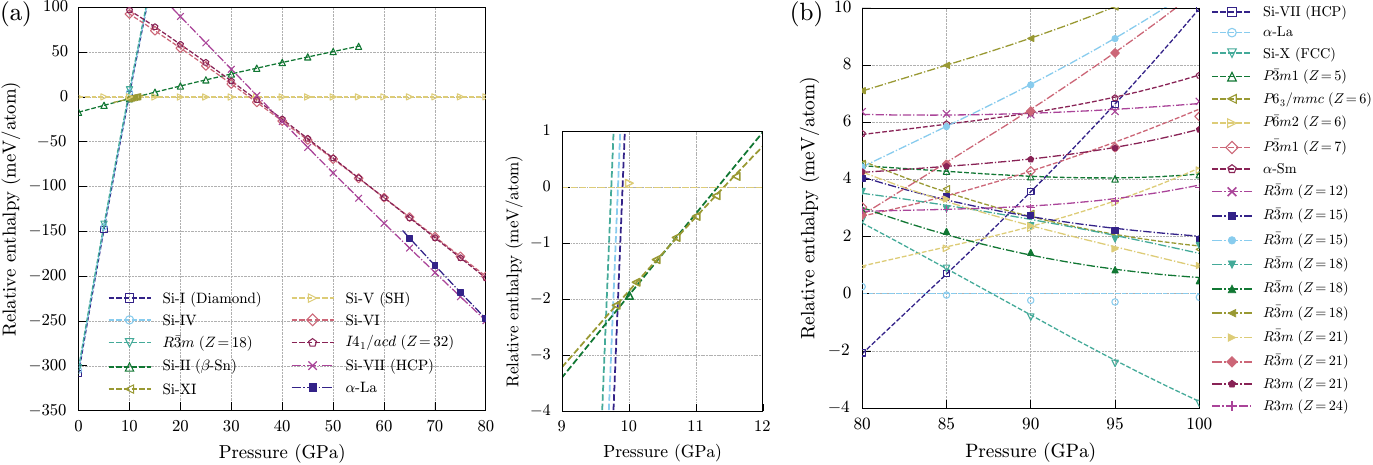}
    \caption{
    Relative enthalpy values computed using DFT calculations for local minimum structures. 
    Local minimum structures with relative enthalpy values exceeding 10 meV/atom across the entire pressure range are excluded from the analysis. 
    The legend indicates the space groups of the structures or their corresponding prototype structures, and \(Z\) denotes the number of atoms in the unit cell.
    (a) Enthalpy values are shown relative to the Si-V structure for pressures from 0 to 80 GPa in the left panel. The relative enthalpies in the central panel are displayed for the 9–12 GPa range, where multiple structures have very similar enthalpy values.
    (b) Enthalpy values are shown relative to the \(\alpha\)-La-type structure for pressures from 80 to 100 GPa.
    }
    \label{fig:stable_plot_Si}
\end{figure*}

Figure \ref{fig:stable_plot_Si} shows the relative enthalpy values from DFT calculations for local minimum structures shown in Fig. 5(b) of the main text.
Structures whose relative enthalpy values exceed 10 meV/atom across the entire pressure range are excluded from Fig. \ref{fig:stable_plot_Si}.
In addition to the stable structures listed in Table I of the main text, three metastable local minimum structures are identified in the 0--80 GPa range in Fig. \ref{fig:stable_plot_Si}(a).
They correspond to the $\text{Si-I\hspace{-1.2pt}V}$ structure, a structure with space group $I4_1/acd$ represented with 32 atoms, and a structure with space group $R\bar3m$.
Moreover, it is also observed that the enthalpy values of the $\text{Si-X\hspace{-1.2pt}I}$ structure are nearly identical to those of the $\beta$-Sn-type structures.
Between 80 and 100 GPa in Fig. \ref{fig:stable_plot_Si}(b), 15 metastable structures with low relative enthalpy values are found.
These structures exhibit different stacking orders of close-packed planes, indicating that elemental Si tends to adopt close-packed structures under high pressures.

\begin{figure}
    \centering
    \includegraphics[width=0.5\linewidth]{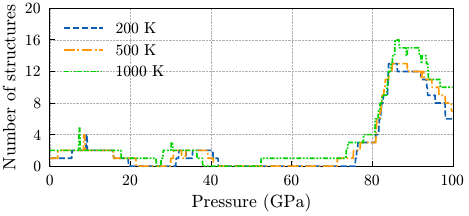}
    \caption{Relationship between pressure and the number of local minimum structures in elemental Si. Only structures with relative Gibbs free energies below 5 meV/atom are included, excluding the global minimum structure. The pressure ranges from 0 to 100 GPa, and the numbers are shown for temperatures of 200, 500, and 1000 K.}
    \label{fig:sscha_meta}
\end{figure}

Figure \ref{fig:sscha_meta} illustrates the relationship between pressure and the number of local minimum structures with relative Gibbs free energies below 5 meV/atom, excluding the global minimum structure, as calculated using the SSCHA with specialized MLPs.
The number of these local minimum structures remains nearly constant as the temperature increases. 
Additionally, many structures exhibit low relative Gibbs free energies in the 80--100 GPa range, where close-packed structures such as the HCP-, $\alpha$-La-, and FCC-type structures are stable. 
The maximum number of local minimum structures with relative Gibbs free energies below 5 meV/atom is 17, observed at 86.5 GPa and 900 K. 
Among these structures, 16 correspond to those with low relative enthalpy, as shown in Fig. \ref{fig:stable_plot_Si}, in the pressure range of 80–100 GPa at 0 K. 

\subsection{Energy errors of polynomial MLPs in SSCHA calculations}

\begin{figure}
    \centering
    \includegraphics[width=0.5\linewidth]{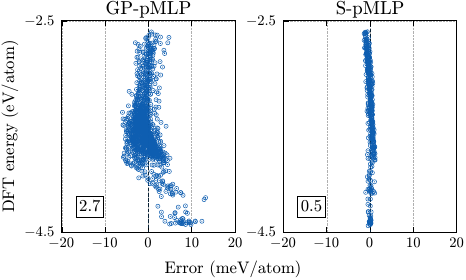}
    \caption{Distributions of energies for sample structures in SSCHA calculations for local minimum structures. The energy and force values are computed using the general-purpose polynomial MLP and the specialized polynomial MLPs, and are compared with those obtained from DFT calculations. The numerical values enclosed in the squares indicate RMSEs for energy, expressed in meV/atom. In the figure, GP-pMLP and S-pMLP stand for general-purpose polynomial MLP and specialized polynomial MLP, respectively.}
    \label{fig:sscha_error}
\end{figure}

Figure \ref{fig:sscha_error} illustrates the distributions of energy errors for representative sample structures obtained from SSCHA calculations of local minimum structures.
Energy values are computed using the general-purpose MLP and the specialized MLP, and are compared against those from DFT calculations.
Although the general-purpose MLP demonstrates reasonable predictive performance in SSCHA calculations, its energy prediction errors tend to be slightly larger for structures with low cohesive energies, which correspond to volumes near equilibrium under low pressure.
In contrast, the specialized MLPs accurately reproduce the DFT energy values across all local minimum structures, thereby enabling more accurate SSCHA calculations.

\subsection{Results under the computational settings of previous studies}

\begin{figure}
    \centering
    \includegraphics[width=\linewidth]{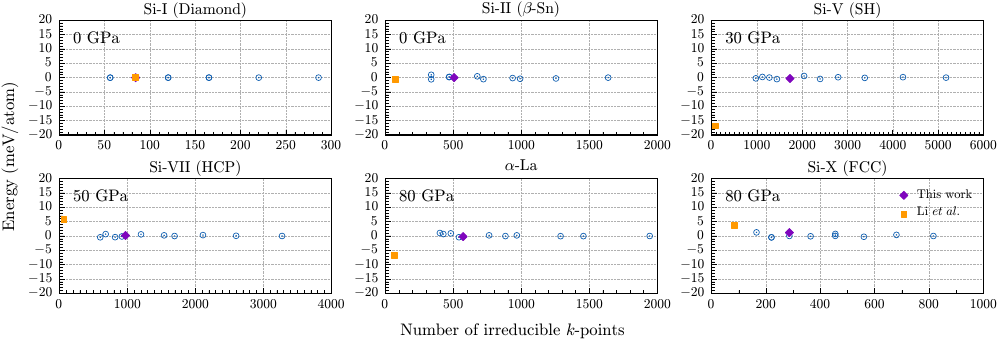}
    \caption{Energy values for six structures calculated using the $k$-point settings adopted by Li $\it{et~al.}$ \cite{Li_JMS_2018} and those employed in the present study. 
Energies are given relative to the values obtained with the densest $k$-point sampling.
The setting used in the present study (0.09 $\mathrm{\AA}^{-1}$) is indicated by purple filled diamonds, whereas that adopted by Li $\it{et~al.}$ is shown by orange filled squares.}
    \label{fig:kpoint}
\end{figure}

Two previous studies have reported phase-diagram calculations for Si under high-pressure and high-temperature conditions \cite{Li_JMS_2018, PhysRevLett.122.125701, PhysRevB.100.144101}.
In this section, we compare the results obtained using the computational settings adopted in previous studies with those obtained using our settings.
Figure \ref{fig:kpoint} presents the energy values for six structures calculated using the $k$-point settings adopted by Li $\it{et~al.}$ \cite{Li_JMS_2018} and those used in the present study.
Moreover, additional DFT calculations were performed with the allowed $k$-point spacing varied from 0.06 to 0.11 $\mathrm{\AA}^{-1}$.
DFT calculations were performed using the \textsc{vasp} code \cite{PhysRevB.47.558, PhysRevB.54.11169, KRESSE199615}.
The cutoff energy was set to 400 eV, and the configuration of the valence electrons in the PAW potential for Si is $3s^{2}3p^{2}$.
These settings are identical to those used by Li $\it{et~al.}$
While the $k$-point settings employed in this study yield converged energies, those used by Li $\it{et~al.}$ do not achieve convergence for the SH-, HCP-, $\alpha$-La-, and FCC-type structures.

\begin{figure}
    \centering
    \includegraphics[width=0.8\linewidth]{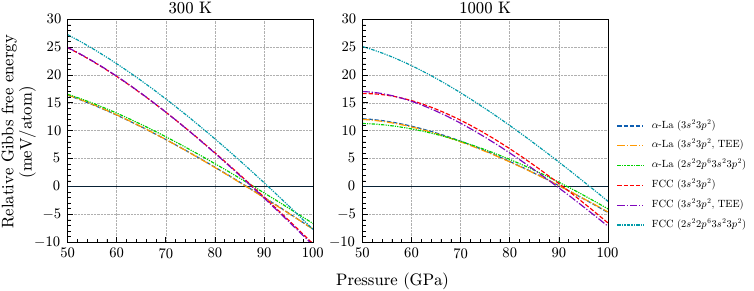}
    \caption{
    Relative Gibbs free energies at 300 K and 1000 K for the $\alpha$-La- and FCC-type structures, computed using DFT-based QHA calculations and referenced to the HCP-type structure. 
    Results are shown for QHA calculations using the PAW potential ($3s^{2}3p^{2}$) without electronic thermal excitations, with electronic thermal excitations, and using the PAW potential ($2s^{2}2p^{6}3s^{2}3p^{2}$) without electronic thermal excitations. 
    In the legend, TEE denotes calculations including the contribution of thermal electronic excitations to the free energy.}
    \label{fig:approx_comparison}
\end{figure}

Paul $\it{et~al.}$ included the contribution of thermal electronic excitations to the free energy \cite{PhysRevLett.122.125701, PhysRevB.100.144101}.
They also employed a PAW potential treating $2s^{2}2p^{6}3s^{2}3p^{2}$ as valence electrons, which differs from the PAW potential used in the present study.
To assess the impact of these choices under high-pressure conditions, we performed additional DFT-based QHA calculations including the same contributions.
The contribution of thermal electronic excitations was evaluated by applying the Fermi–Dirac distribution at the target temperature to the one-electron states obtained from single-point DFT calculations.
The cutoff energy was set to 1100 eV, and the allowed spacing between $k$ points was approximately 0.09 $\mathrm{\AA}^{-1}$.
Because of the high computational cost associated with this PAW potential, the supercells were limited to 32 atoms and were constructed by expanding the conventional unit cells of the respective structures.

Figure \ref{fig:approx_comparison} shows the relative Gibbs free energies of the $\alpha$-La- and FCC-type structures, referenced to the HCP-type structure. 
As shown in Fig. \ref{fig:approx_comparison}, the contribution of thermal electronic excitations is negligible. 
The effect of the PAW potential is also negligible for the $\alpha$-La-type structure, whereas for the FCC-type structure it remains small, on the order of a few meV/atom.

\bibliography{references}